\documentclass[aps,prb,twocolumn,groupedaddress,notitlepage]{revtex4-1}

\usepackage{graphicx,amsmath,color,amsfonts,bm}

\newcommand{\be}{\begin{equation}}
\newcommand{\ee}{\end{equation}}
\newcommand{\bea}{\begin{eqnarray}}
\newcommand{\eea}{\end{eqnarray}}

\begin{document}

\title{Probing one-dimensional systems via noise magnetometry with single spin qubits}

\author{Joaquin~F. Rodriguez-Nieva$^{1}$, Kartiek Agarwal$^{2}$, Thierry Giamarchi$^{3}$, Bertrand~I. Halperin$^{1}$, Mikhail~D. Lukin$^{1}$, Eugene Demler$^{1}$}
\affiliation{$^1$Department of Physics, Harvard University, Cambridge, MA 02138, USA}
\affiliation{$^2$Department of Electrical Engineering, Princeton University, Princeton NJ 08544, USA}
\affiliation{$^3$Department of Quantum Matter Physics, University of Geneva, Geneva 1211, Switzerland}

\date{\today}
\begin{abstract}

The study of exotic one-dimensional states, particularly those at the edges of topological materials, demand new experimental probes that can access the interplay between charge and spin degrees of freedom. One potential approach is to use a single spin probe, such as a Nitrogen Vacancy center in diamond, which has recently emerged as a versatile tool to probe nanoscale systems in a non-invasive fashion. Here we present a theory describing how noise magnetometry with spin probes can directly address several questions that have emerged in experimental studies of 1D systems, including those in topological materials. We show that by controlling the spin degree of freedom of the probe, it is possible to measure locally and independently local charge and spin correlations of 1D systems. Visualization of 1D edge states, as well as sampling correlations with wavevector resolution can be achieved by tuning the probe-to-sample distance. Furthermore, temperature-dependent measurements of magnetic noise can clearly delineate the dominant scattering mechanism (impurities vs. interactions) -- this is of particular relevance to quantum spin Hall measurements where conductance quantization is not perfect. The possibility to probe both charge and spin excitations in a wide range of length scales opens new pathways to bridging the large gap between atomic scale resolution of scanning probes and global transport measurements. 
\end{abstract}



\maketitle

\tableofcontents

\section{Introduction}

One-dimensional phases of matter exhibit a myriad of exotic phenomena including non-Fermi liquid behavior, charge-spin separation, and power-law scaling of charge and spin correlations.\cite{1995bosonizationreview,1998bosonizationreview,giamarchibook} Reinvigorated interest in such phases resulted from the recent realization of 1D edge states emerging in topological materials, for instance quantum spin Hall states.\cite{2004qshsemiconductor,2005PRLqshobservation,2005qshimaging,2007qsh-exp} The design of new experimental probes to access these interesting and exotic states is highly desirable but equally demanding. For instance, because in many cases 1D states live at the edges of higher dimensional systems, using experimental probes that require macroscopic samples to obtain a measurable signal, e.g. neutron or light scattering, is very challenging. Furthermore, probes that can bridge the large lengthscale gap between atomic scale resolution of scanning tunneling probes and global transport measurement are on high demand, in particular to obtain correlations with wavevector resolution. Accessing physics at the nanometer scale, however, impose stringent requirements on probe size. 

Motivated by the rapid progress in magnetic noise spectroscopy with single spin qubits, such as Nitrogen Vacancy (NV) centers in diamond,\cite{2014nvreview,2008nvlukin,2015kolkowitz,2015nvmagnons,2013mamin} here we outline pathways to exploit single spin probes to access 1D physics in a broad range of 1D systems, including those emerging in topological materials. Spin probes harness the fluctuating magnetic field induced by quantum and thermal fluctuations of 1D charged and spin modes. By measuring the spin relaxation time $T_1$ as a function of experimentally tunable parameters, e.g. temperature ($T$), probe-to-sample distance ($R$) and spin probe polarization, 1D correlations can be obtained. There are several key advantages of single spin probes. Because of their atomic size, spin probes enable measurements with nanometer resolution, much smaller than the micron scales achievable via NMR.\cite{2002nmrreview} This feature also grants access to spin fluctuations, which can only be detected at nanometer scale proximity due to short range dipole-dipole interaction, and makes the measurement insensitive to boundary effects, such as the contacts. In addition, because the electromagnetic coupling between the probe and the sample decays as a power law, different from the exponential decay of scanning tunneling currents, single spin probes can access a broader range of lengthscales, from few to hundred nanometers. Another interesting feature is that, because spin probes do not require driving fields, i.e. they are driven by charge and spin fluctuations in the sample, they are minimally invasive. 

\begin{figure}
  \centering\includegraphics[scale=1.0]{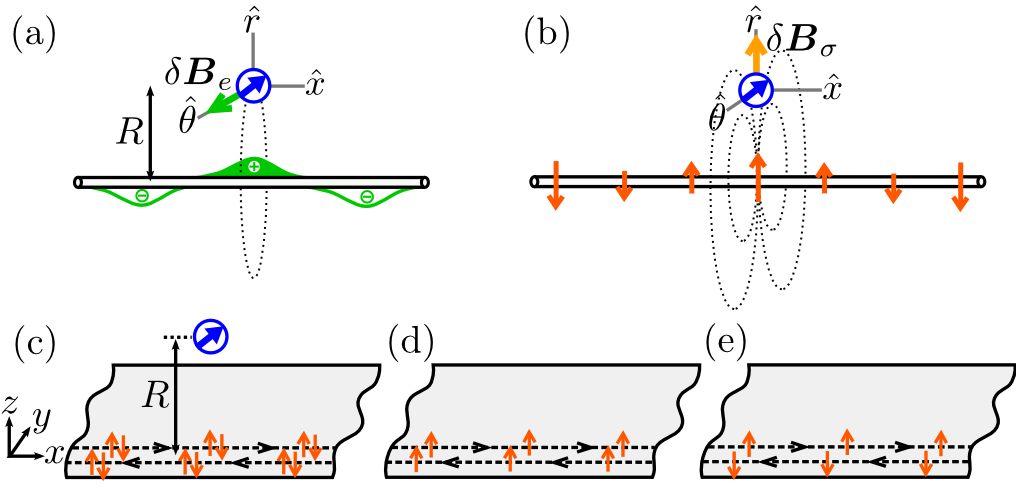}
  \caption{Separating charge and spin fluctuations of 1D systems using single spin probes. For 1D systems, the spin probe (blue spin) can {\it independently} probe (a) charge fluctuations and (b) spin fluctuations in the quantum wire. Charge fluctuations induce magnetic noise $\delta{\bm B}_e$ in the azimuthal direction ($\hat\theta$), whereas spin fluctuations induce magnetic noise $\delta{\bm B}_\sigma$ primarily in the radial ($\hat{r}$) and azimuthal ($\hat\theta$) directions. The spin degree of freedom of the probe can be used to filter the charge fluctuations from the spin fluctuations. To understand how the internal structure of 1D states affects noise, here we use three prototypical examples: (c) the most typical case of two counterpropagating channels with an SU(2) degree of freedom; (d) a spin-polarized edge state comprised of right and left movers with equal spin polarization; (e) a non-chiral helical edge states comprised of right and left movers with opposite polarizations. }
  \label{fig:schematics}
\end{figure}

The success of this technique in the study of excitations in higher dimensional materials, such as metallic surfaces\cite{2015kolkowitz} and ferromagnets,\cite{2015nvmagnons,2013mamin,2017magnonmu} combined with theoretical proposals to measure electron viscocity in the hydrodynamic regime,\cite{2017kartieknoise} forecasts grand new vistas in 1D. Interestingly, as compared to larger dimensional systems, we find that the noise behavior in 1D systems features two fundamental differences, which can be rendered into practical advantages. First, it was shown that magnetic fluctuations emerging from metallic surfaces are dominated by transverse charge currents, thus making noise originating from longitudinal currents (i.e., charged modes) and spin fluctuations inaccessible.\cite{2017kartieknoise} The absence of transverse charge currents in 1D grants access to charge and spin-induced fluctuations simultaneously when the probe is sufficiently close to the sample. Second, not only is it possible to access both charge and spin excitations but also {\it separate} them, even when they are comparable. As shown in Fig.\ref{fig:schematics}(a)-(b), this feature arises because of the spin degree of freedom of the probe which allows to measure $B$-field in different directions: whereas charge fluctuations induce magnetic fields in the azimuthal direction $\hat\theta$, spin fluctuations induce magnetic fields in all spatial directions (the radial and azimuthal components dominate at long wavelengths, see below). Since the relaxation time of a spin probe is determined only by magnetic field fluctuations perpendicular to the orientation of the spin probe, we see that charge-induced noise can be filtered from spin-induced noise by aligning the spin probe with the azimuthal direction, i.e. the probe acting as a vector magnetometer.\cite{2010vectormagnetometer} Therefore, spin probes represent an important departure, for instance, from scanning tunneling probes which cannot separate charge and spin excitations.\cite{1996chargespinseparation,1998chargespinseparation,1999chargespinseparation,2005chargespinseparation,2009chargespinseparation} For example, even if the tunneling tip is spin-polarized, this technique still requires tunneling of electrons into the sample and, as such, does not work in insulating materials with spin excitations. 

Besides polarization direction, other experimentally tunable parameters are available to access different features of 1D states, such as scattering and transport. For instance, by tuning the probe-to-sample distance $R$ and scanning noise at different length scales, it is possible to sample correlations with wavevector resolution and diagnose transport behavior, i.e. whether charge and spin density waves propagate  ballistically or are pinned by disorder. Furthermore, we find that scattering is the key factor leading to non-universal power law behavior of noise ${\it vs}$ $T$; the specific $T$-dependence hints at the nature of backscattering, i.e. whether it is single-particle or interaction-assisted. 

Turning our discussion to specific 1D models, we first note that, at the sub-THz frequencies characteristic of spin probes in current experimental setups, a good starting point to describe general 1D systems is the Luttinger Liquid (LL) theory.\cite{1965luttinger,1965ll-mattis,1981haldane} To capture the key aspects of magnetic noise measurents, we exploit minimal models that qualitatively describe the effects of scattering, interactions, and internal structure of 1D states. Because we need at least two 1D channels to describe scattering, here we mainly focus on non-chiral, two-channel systems such that one channel is right-moving and the other is left-moving. 

To aggregate the internal structure of 1D states into our discussion, we consider three minimal models. First, we consider that each channel has an SU(2) degree of freedom [Fig.\ref{fig:schematics}(c)], which is the most usual case describing quantum wires or metallic nanotubes. Second, we consider a spin-polarized LL in which excitations are comprised of left-moving and right-moving modes with equal spin polarization [Fig.\ref{fig:schematics}(d)]. Third, and motivated by the recent realization quantum spin Hall states, we consider a pair of counterpropagating helical edge state. Contrary to the previous case, the right-moving and left-moving excitations have opposite spin polarizations [Fig.\ref{fig:schematics}(e)]. The helical state differs from the SU(2) and spin-polarized states in several important ways. In particular, when time-reversal symmetry is present, carriers cannot be backscattered by disorder as this would require a spin flip;\cite{2005kanemele} as such, backscattering needs to be assisted by interactions.\cite{2009qshkondo,2010qshrashba,2011qshimpurity,2012qshbackscattering,2013helicalresistance,2014helicalglazman} Below we describe how the interplay between scattering, interactions and internal structure of carriers affect the noise spectrum. 

The outline of the present work is as follows. In Sec.\ref{sec:measurement}, we present the theory of magnetic noise spectroscopy and, in particular, how $T_1$ can be computed from charge and spin density correlations of the 1D system. In Sec.\ref{sec:noisell}, we focus on the magnetic noise behavior of clean wires described within the LL theory, and describe how it varies as a function of experimentally tunable parameters, in particular probe-to-sample distance, temperature and interactions strength. In Sec.\ref{sec:noiseweakdisorder}, we introduce weak, dense disorder (Gaussian disorder) and describe how the noise behavior is qualitatively modified from the LL behavior. In Sec.\ref{sec:noisestrongdisorder}, we introduce sparse, strong impurities (Poisson disorder) and describe the effect on magnetic noise. In Sec.\ref{sec:extension}, we discuss scenarios that go beyond our minimal two-channel model and, in Sec.\ref{sec:conclusion}, we summarize the main results. 

\section{Relaxation time measurement: general formalism}
\label{sec:measurement}

We begin by describing a general formalism that allows to relate the relaxation time $T_1$ to charge and spin density correlations of general 1D systems. With this objective in mind, we first consider a generic 1D system with charge and spin dynamics governed by the action ${\cal S}_{\rm 1D}$ (below we introduce specific microscopic models). The coupled dynamics of the wire and the electromagnetic field is described by the action
\be
{\cal S} ={\cal S}_{\rm 1D} - \iint dt d{\bm r} \left[ F^{\mu\nu}F_{\mu\nu} /4\mu_0 + A_\mu J^\mu \right],
\label{eq:action}
\ee
where we use standard 4-vector covariant notation, $F^{\mu\nu}$ denotes $F^{\mu\nu} = \partial^{\mu}A^\nu - \partial^\nu A^\mu$, $A^\mu $ is the vector potential, $\mu_0$ is the vacuum permeability, and coordinates are in three-dimensional space, ${\bm r} = (x,y,z)$. Assuming the wire to be at ${\bm r} = (x,0,0)$, charge and spin density fluctuations in the wire act as sources of electromagnetic field via the term $J^\mu = J_{e}^\mu + J_{\sigma}^\mu$: 
\be
\begin{array}{c}
\displaystyle J_e^\mu = e\left[c\rho_e\left(x,t\right) , 0 , 0 , j_e\left(x,t\right)\right]\delta(y)\delta(z), \\ \\ 
\displaystyle J_\sigma^\mu = (0,\nabla\times{\bm m}), \quad{\bm m} = g_\sigma \mu_{\rm B} \sum_{i=x,y,z}\rho_i(x,t)\delta(y)\delta(z)\hat{\bm e}_i, 
\end{array}
\label{eq:sources}
\ee
with $c$ is the speed of light, $\mu_{\rm B}$ the Bohr magneton, and $g_{\sigma}$ the $g$-factor of the spin modes in the wire. 
For a 1D system, the carrier density, $\rho_e $, and the current, $j_e$, are related by the continuity equation, $\partial_t\rho_e = -\partial_x j_e$. Furthermore, for systems with an SU(2) degree of freedom, any of the spin components $\rho_{x,y,z}(x,t)$ can fluctuate; for systems with polarized spin in direction $\hat{\bm n}$, fluctuations are given by $\rho_i = \rho_{\sigma} \hat{\bm n}_i$. 

In thermal equilibrium, Eq.(\ref{eq:action}) combined with the sources in Eq.(\ref{eq:sources}) give rise to fluctuations in electric and magnetic field induced by $\rho_e$ and $\rho_\sigma$ in the quantum wire, as well as vacuum electromagnetic fluctuations. A spin probe at position ${\bm r}$ is sensitive to fluctuations in magnetic field $\delta {\bm B}({\bm r},t)$.\cite{1975agarwal,19993dnoise,2007noisereview,2012langsjoen,2013poudel} For concreteness, here we consider a spin-1/2 probe with an intrinsic level splitting $\hbar\omega$. The spin dynamics is governed by the time-dependent Hamiltonian ${\cal H}_{\rm spin} = (\hbar \omega/2)\hat{\bm n}_{\rm p}\cdot{\boldsymbol\sigma} + g_{\rm s}\mu_{\rm B}[{\boldsymbol\sigma}\cdot{\delta{\bm B}({\bm r},t)}]$, where $\hat{\bm n}_{\rm p}$ is the direction of the intrinsic polarizing field. If $\hat{\bm n}_{\rm p} = \hat{\bm z}$, for instance, the relaxation time can be calculated using Fermi Golden's rule, which yields $1/T_1 = [{\cal N}_{xx}(\omega) + {\cal N}_{yy}(\omega) - 2{\rm Im}[{\cal N}_{xy}(\omega)]]$. The tensor ${\cal N}_{ij}({\bm r},\omega)$, 
\be
{\cal N}_{ij} ({\bm r},\omega) = \frac{(g_{\rm s}\mu_{\rm B})^2}{2\hbar^2} \int_{-\infty}^{\infty} dt \langle \{ \delta B_i({\bm r},t) , \delta B_j({\bm r},0) \} \rangle e^{i\omega t},
\label{eq:noise}
\ee
quantifies the amplitude of magnetic fluctuations, i.e. the magnetic noise in units of ${\rm sec}^{-1}$, at the position of the probe [see derivation in Appendix\,\ref{app:relaxationtime}]. Importantly, a spin probe with level splitting $\hbar\omega$ only couples to magnetic modes oscillating at frequency $\omega$. In Eq.(\ref{eq:noise}), $\{,\}$ denotes anticommutation and $\langle.\rangle$ denotes statistical average on the canonical ensemble at temperature $T$. 

The calculation of the magnetic noise, Eq.(\ref{eq:noise}), for a generic 1D system can be simplified under several legitimate assumptions. First, we assume translational invariance in the direcion of the wire ($\hat{\bm x}$), which is descriptive of the long wavelength behavior expected to occur at the characteristic sub-THz frequencies. Translational-symmetry breaking effects, e.g. disorder or commensurability, can be accounted for in terms of self-energy corrections, as will be described below. Second, we assume quasistatic dynamics of the electromagnetic field, such that $\delta B_{i}$ tracks $\rho_e$ and $\rho_\sigma$ without any retardation effects; this is generally valid in solid state systems because excitations propagate with velocities much smaller than $c$. Under these assumptions, the 1D charge density, $\rho_e$, and the spin densities, $\rho_{x,y,z}$, give rise to four orthogonal electromagnetic modes:
\be
\delta {\bm B}({\bm r},t) = \frac{1}{\sqrt{L}}\sum_{q\omega m}{\bm H}_{m}(q,y,z,\omega)e^{i(qx-\omega t)}\rho_m(q,\omega),
\label{eq:bfield}
\ee 
where ${\bm H}_{m}$ is the magnetic eigenfunction associated with each mode $m = e,x,y,z$, and $L$ is the length of the wire. Vacuum electromagnetic fluctuations also contribute to Eq.(\ref{eq:bfield}), but we expect these to be negligly small compared to wire-induced fluctuations. 

The solution of Maxwell's equation for charge and spin fluctuations is discussed in Appendix\,\ref{app:emmodes}. To illustrate the qualitative behavior, here we present a more intuitive approach in the simple geometry of Fig.\ref{fig:schematics}(c) in which the probe is at ${\bm r} = (0,0,R)$. Focusing first on charge fluctuations and assuming quasistatic behavior, we can calculate magnetic field via Biot-Savart's law, $\delta {\bm B}_e(t) = \frac{\mu_0 e}{4\pi}\int_{-\infty}^{\infty} dx' j_e(x',t) \frac{\hat{\bm x}\times({\bm r}'-{\bm r})}{|{\bm r}'-{\bm r}|^{3}}$, with ${\bm r}' = (x',0,0)$ wire coordinates. Currents can be related to charge density via the continuity equation $\partial_t \rho =  - \partial_x j$, i.e. $j(x,t) = (\omega/q)\rho_m(q,\omega)e^{i(qx-\omega t)}/\sqrt{L}$, so that $\delta{\bm B}_e(t)$ can be rewritten as $\delta {\bm B}_e(t) = \frac{\mu_0 e \omega \rho_e(q,\omega)}{4\pi\sqrt{L}}\int_{-\infty}^{\infty} dx' \frac{R e^{iqx'} }{q(R^2 + x'^2)^{3/2}}\hat{\bm y}$ [here we defined $\rho(x) = \frac{1}{\sqrt{L}}\sum_q \rho(q)e^{iqx}$]. Integration in $x'$ leads to magnetic field written in the form (\ref{eq:bfield}) with 
\be
{\bm H}_{e}(q,0,R,\omega)= -\frac{\mu_0e}{2\pi}\left( \begin{array}{c} 0 \\ \omega K_1(q R) \\ 0 \end{array} \right), 
\label{eq:hfielde}
\ee
where $K_n(x)$ is the $n$-th modified Bessel function of the second kind. Naturally, $\delta{\bm B}_e$ points in the azimuthal direction. We also note that $K_n(x)$ behaves as a power law for $x\lesssim 1$, $K_n(x) \propto 1/x^n$, but decays exponentially for $x \gtrsim 1$, $K_n(x) \propto e^{-x}/\sqrt{x}$. Such transition at $qR \approx 1$ occurs because for large $R$ there is a negligible signal due to wave interference of the electromagnetic field. 

Contrary to charge density, the spin-induced electromagnetic field has components in all three spatial directions, but the radial and axial components (with respect to the axis of the wire) dominate in the long-wavelength limit, $qR \ll 1$. For instance, assuming that the spin is polarized in the $\hat{\bm z}$ direction transverse to the wire, we find 
\be
\begin{array}{rl}
\displaystyle{\bm H}_\sigma (q,0,R,\omega) = &\displaystyle \frac{g_\sigma\mu_0\mu_{\rm B}}{4\pi} 
\\
&\displaystyle\times\left( \begin{array}{c} -2iq^2K_1(qR) \\ 0 \\ q^2\left[K_0(qR)-K_2(qR)\right] \end{array}\right).
\end{array}
\label{eq:hfields}
\ee
As such, $\delta {\bm B}_e$ and $\delta {\bm B}_\sigma$ can be separated by exploiting the probe polarization direction. The spin-induced field exhibits the same power-law to exponential transition occurring at $qR \approx 1$ as in the charge-induced field. 

After a series of uneventful steps described in Appendix\,\ref{app:effectiveaction}, in particular replacing Eq.(\ref{eq:bfield}) into Eq.(\ref{eq:noise}), and expressing density fluctuation in terms of dissipation in the wire, we find charge and spin noise given by
\be
\begin{array}{r}
\displaystyle{\cal N}_m({\bm r},\omega) = \left(\frac{g_{\rm s}\mu_{\rm B}}{\hbar}\right)^2{\rm coth}\left(\frac{\hbar\omega}{2k_{\rm B}T}\right)\frac{1}{L}\sum_q F_{m}(q,y,z,\omega) \\ 
\displaystyle\times{\rm Im}\left[{\cal C}_{\rho_m\rho_m}^{\rm R}(q,\omega)
\right], \quad\quad m=e,\sigma. 
\label{eq:noise2}
\end{array}
\ee
Here ${\cal C}_{\rho_m\rho_m}^{\rm R}(q,\omega)$ is the short-hand notation for the retarded density-density correlation function, ${\cal C}_{AB}^{\rm R}(q,\omega) = - i \int_{-\infty}^{\infty} dt \Theta(t) \langle \left[ A(t),B(0)\right]\rangle e^{i\omega t}$. For spin noise we aggregate the three components of spin fluctuations $\rho_{x,y,z}$ into a single term ${\cal N}_\sigma$. The factors $F_m(q,y,z,\omega)$ quantify the electromagnetic coupling between the wire and the probe: $F_e(q,y,z,\omega) = |{\bm H}_{e}(q,y,z,\omega)|^2$, for charge modes, $F_\sigma(q,y,z,\omega) = \sum_m|{\bm H}_{m}(q,y,z,\omega)|^2$ for SU(2) spin modes (fluctuations in each spin component are independent) and $F_\sigma(q,y,z,\omega) = |\sum_m\hat{\bm n}_m{\bm H}_{m}(q,y,z,\omega)|^2$ for spin-polarized states. We note that Eq.(\ref{eq:noise2}) resembles the standard $1/T_1$ equation for NMR relaxation, except for the $q$ and ${\bm r}$ dependent form factors. Further, in the case of charge fluctuations, noise measurements can be related to conductivity $\sigma(q,\omega)$ measurements at finite $q$ and $\omega$. In particular, by using the continuity equation $\omega\rho_e = q j_e$ and the definition $\sigma(q,\omega) = \langle j_q\bar{j}_q\rangle /i\omega$, Eq.(\ref{eq:noise2}) can be expressed as ${\cal N}_e \propto \sum_q q^2 F(q,y,z,\omega) {\rm Re}[\sigma(q,\omega)]/\omega$.

Equation (\ref{eq:noise2}) captures the essense of the noise measurement by making the connection between $T_1$ and charge and spin density correlations in a generic 1D system. In particular, by tuning $R = \sqrt{y^2+z^2}$, it is possible to sample fluctuations at different wavevectors $q$ by changing the weight of the form factor $F_{m}(q,y,z,\omega)$. For instance, for $ q \lesssim 1/R$, the form factors as a function of $q$ behaves as $[q^2F_e(q)] \sim 1$ for charge noise, and $[q^2F_\sigma(q)]\sim q^2$ for spin noise, i.e. there is finite sampling of charge and spin fluctuations for all modes with wavevectors $q\lesssim 1/R$. For $q\gtrsim 1/R$, the form factor for charge noise behaves as $[q^2F_e(q)] \sim qe^{-2qR}$, and for spin noise as $[q^2F_\sigma(q)]\sim q^5e^{-2qR}$. As such, there is a sharp cutoff in the sampling of fluctuations occurring at $q\sim 1/R$ introduced by the exponential $q$ dependence of $F_m(q,y,z,\omega)$. Such wavevector selectivity allows to study correlations with wavevector resolution and which, as we will see, is a useful feature in the study of disordered systems. 

We now proceed to specify microscopic 1D models from which the density correlation ${\cal C}_{\rho_m\rho_m}^{\rm R}(q,\omega)$ can be computed explicity. This is the objective of the next two sections. 

\section{Noise from Luttinger liquids}
\label{sec:noisell}

To capture the microscopics of the wire, we use the bosonization description for 1D electronic systems.\cite{giamarchibook} This framework is ideally suited for our purposes given the typically small sub-THz probing frequencies, much smaller than typical bandwidths in electronic systems, and its ability to describe 1D states of different flavors. Further, it provides a good starting point to describe more complex scenarios such as disordered wires. We set the stage by discussing magnetic noise in ballistic 1D channels with an SU(2) degree of freedom. Afterwards, we describe noise in clean spin-polarized and helical channels and point out the differences with the SU(2) case. 

\subsection{Case I: SU(2) channels}

The motion of spinful fermions in a 1D channel can be described with a bosonic 1D action with separated charge and spin degrees of freedom: 
\be
{\cal S}_{\rm 1D} = \iint dt dx \sum_{m=e,\sigma} \left[\,i\Pi_m \partial_t \phi_m - {\cal H}_{m}(\phi_m,\Pi_m)\right].
\label{eq:action1d}
\ee
The bosonized degrees of freedom, $\Pi_m$ and $\phi_m$, are canonically conjugate, $[\phi_m(x),\Pi_{m'}(x')] = i\delta_{mm'} \delta(x-x')$, and describe charge ($m=e$) and spin ($m=\sigma$) excitations. In the absence of scattering, dynamics is governed by a quadratic Hamiltonian of the form
\be
{\cal H}_{m}(\phi_m,\Pi_m) = \frac{\hbar v_{\rm F}}{2\pi} \left[(\pi {\Pi}_m)^2 + (\partial_x\phi_m)^2\right],\quad m=e,\sigma,
\label{eq:h1dspinless}
\ee
where $v_{\rm F}$ is the Fermi velocity. In the long wavelength limit, charge and spin density are related to the bosonic degrees of freedom via $\rho_{e,\sigma} = - \sqrt{2}\partial_x\phi_{e,\sigma}/\pi$. This linear mapping $(\rho_e,\rho_\sigma)\leftrightarrow(\phi_m,\Pi_m)$ is valid up to spatially oscillating terms with wavevector $k_{\rm F}$, the Fermi wavector. Because these rapidly oscillating density terms produce negligibly small evanescent magnetic fields at distances larger than a few atomic sites, we do not explicitly keep track of them. Further, because of SU(2) symmetry, spin fluctuations in all spatial directions are equal, ${\cal C}_{\rho_m\rho_m}^{\rm R}(q,\omega) = {\cal C}_{\rho_\sigma\rho_\sigma}^{\rm R}(q,\omega)$. Importantly, we note that Eq.(\ref{eq:h1dspinless}) does not include Coulomb interactions. This avoids double-counting the Coulomb potential which is mediated by $A^{\mu}({\bm r},t)$ already included in the full action in Eq.(\ref{eq:action}). 

Having established the microscopic model via Eq.(\ref{eq:action}) and Eqs.(\ref{eq:action1d})-(\ref{eq:h1dspinless}), and the mapping $(\rho_e,\rho_\sigma) \leftrightarrow (\phi_m,\Pi_m)$, we now proceed to calculate ${\cal C}_{\rho_m\rho_m}^{\rm R}(q,\omega)$ in Eq.(\ref{eq:noise2}). The linear nature of the mapping $(\rho_e,\rho_\sigma) \leftrightarrow (\phi_m,\Pi_m)$ simplifies calculations significantly. First, because $A^\mu({\bm r},t)$ couples linearly to $\phi$ and $\Pi$ in Eq.(\ref{eq:action}), and the action is quadratic in $A^\mu({\bm r},t)$, we can integrate exactly the electromagnetic modes coupled to charge/spin densities and incorporate them into an effective Hamiltonian $\bar{\cal H}_{m}$ with renormalized parameters. Secondly, because of the linear mapping $(\rho_e,\rho_\sigma)\leftrightarrow(\phi_m,\Pi_m)$, charge and spin density correlators are two-point correlation functions in the bosonic field, which are straight-forward to calculate for quadratic Hamiltonians. 

Using these two simplifications, we proceed to obtain ${\cal C}_{\rho_m\rho_m}^{\rm R}(q,\omega)$. Because of translational invariance in the $\hat{\bm x}$ direction, it is convenient to rewrite the bosonic fields in Fourier space, $\phi_m(x) = \frac{1}{\sqrt{L}}\sum_q \phi_m(q)e^{iqx}$ and $\Pi_m(x) = \frac{1}{\sqrt{L}}\sum_q \Pi_m(q)e^{iqx}$. Spatial integration of the electromagnetic degrees of freedom result in the effective Hamiltonian 
\be 
\bar{\cal H}_{m}(\phi_m,\Pi_m) = \frac{\hbar v_m}{2\pi} \left[ {\cal K}_m [\pi\Pi_m(q)]^2 + q^2 \phi_m^2(q)/ {\cal K}_m\right],
\label{eq:effectivehamiltonian}
\ee
where the parameters $v_m$ and ${\cal K}_m$ are the renormalized velocity and the Luttinger parameter in the charge ($m=e$) and spin sectors ($m=\sigma$). The value of ${\cal K}_m$ quantifies the charge ($m=e$) and spin ($m=\sigma$) compressibility. In particular, at small $q$, renormalization of the Luttinger parameters in the charge sector is governed by Coulomb energy induced by the charge density ($=\frac{\epsilon_0}{2}\int d{\bm r}|\delta{\bm E}_e({\bm r},t)|^2$), which leads to a ${\rm ln}(1/qr_*)$ dependence:
\be
\begin{array}{c}
\displaystyle v_e = v_{\rm F}\sqrt{1 + \delta_e },\quad {\cal K}_e =1/ \sqrt{1+\delta_e},\\ \\ 
\displaystyle \delta_e = (e^2/4\pi\varepsilon_0\hbar v_{\rm F})\left[{\rm ln}(2/qr_*) - \gamma\right],
\label{eq:luttingerparameters}
\end{array}
\ee
where $\gamma = 0.57721\ldots$ is the Euler constant, $r_*$ is the effective radius of the wire, and $\varepsilon_0$ is the vacuum permittivity (see details in Appendix\,\ref{app:effectiveaction} or Chapter 4 of Ref.[\onlinecite{giamarchibook}]).  We also note that both $v_e$ and ${\cal K}_e$ are $q$-dependent. Integration of the electromagnetic field induced by spin modes leads to negligible corrections of the Luttinger parameters on the order of $\delta_\sigma \approx \mu_0\mu_{\rm B}^2  / r_*^2 \hbar v_{\rm F}\sim 10^{-5}$, where we used $r_*\sim 1\,{\rm nm}$ and $v_{\rm F} \sim 10^4\,{\rm m/s}$. As such, in the spin sector we use 
\be
{\cal K}_\sigma = 1,\quad v_\sigma = v_{\rm F}.
\ee

For the quadratic Hamiltonian in Eq.(\ref{eq:effectivehamiltonian}), calculation of ${\cal C}_{\rho_m\rho_m}^{\rm R}(q,\omega)$ is straight-forward:
\be
{\rm SU(2):}\quad\quad{\cal C}_{\rho_m\rho_m}^{\rm R}(q,\omega) = \frac{2}{\pi} \frac{{\cal K}_m v_m q^2 }{(v_m q)^2 -(\omega+i\epsilon)^2 }, 
\label{eq:rhorho}
\ee
where $\epsilon$ is an infinitesimal positive constant. 

Using ${\cal C}_{\rho_m\rho_m}^{\rm R}(q,\omega)$ in Eq.(\ref{eq:noise2}), we find charge and spin noise in a LL given by 
\be
\begin{array}{rl}
\displaystyle{\cal N}_m (\omega,T,R)= & \displaystyle\frac{(g_{\rm s}\mu_{\rm B})^2}{8\pi} \int_0^\infty \frac{dq}{2\pi} F_m(q,y,z,\omega) \\ & \displaystyle \times  {\rm Im}\left[\frac{{\rm coth}(\hbar\omega / 2k_{\rm B} T){\cal K}_m v_m q^2}{(v_mq)^2 - (\omega + i\epsilon)^2}\right].
\end{array}
\label{eq:noisell}
\ee
Equation (\ref{eq:noisell}) summarizes the essence of magnetic noise measurements in LLs and the key dependencies as a function of experimentally tunable parameters, namely polarization direction, $R$, and $T$ (for the purposes of current experimental setups, we take $\omega$ as fixed). In particular, the probe samples charge and spin fluctuations at all $q$ wavevectors, but only picks those modes which resonate with the spin probe frequency $\omega$. This feature is manifested by the $\delta$-function in the integrand of Eq.(\ref{eq:noisell}) introduced by ${\rm Im}[1/[(v_m q)^2 - (\omega+i\epsilon)^2]]\approx \delta(v q - \omega)/\omega$. Encoded in Eq.(\ref{eq:noisell}) is also the ability to measure independently charge and spin noise, which is possible due to the spin degree of freedom of the probe. 

To give a gauge of $T_1$ values encountered in experiments, we evaluate Eq.(\ref{eq:noisell}) in the regime $qR\lesssim 1$, such that Eqs.(\ref{eq:hfielde}) and (\ref{eq:hfields}) can be replaced by their asymptotic values, and $\hbar\omega\lesssim k_{\rm B}T$ such that ${\rm coth}(\hbar\omega / 2 k_{\rm B}T) \approx 2 k_{\rm B}T / \hbar \omega$. This results in 
\be
\begin{array}{c}
\displaystyle\frac{1}{T_{1,e}} = \frac{(\mu_0\mu_{\rm B} e)^2}{(2\pi)^2\hbar^3}\frac{g_{\rm s}^2 k_{\rm B}T}{R^2}{\cal K}_e,\\
\displaystyle\frac{1}{T_{1,z}} = \frac{(\mu_0\mu_{\rm B}^2)^2}{(4\pi)^2\hbar^3}\frac{g_{\rm s}^2k_{\rm B}T}{R^4}\frac{g_{\sigma}^2}{v_{\rm F}^2},
\end{array}
\label{eq:estimates}
\ee
where ${\cal K}_e$ is evaluated at $q=\omega/v_e$, see Eq.(\ref{eq:luttingerparameters}). The first factor of $1/T_{1,m}$ is a combination of universal constants reflecting the coupling of charge modes with the spin probe in $T_{1,e}$, and spin-spin coupling in the case of $T_{1,z}$; the second factor contains experimental parameters, namely the $g$-factor of the probe $g_{\rm s}$, the probe-to-sample distance $R$, and temperature ($T$); the third factor contains 1D system parameters. The relation $1/T_1 \propto g_{\rm s}^2k_{\rm}T / \hbar $ resembles the Korringa law\cite{1950Korringa} apart from geometrical factors which arise because the spin probe is not in the system's bulk.

For estimates, we use $g_{\rm s}= g_{\sigma} = 1$, $v_{\rm F} \sim 10^4\,{\rm m/s}$, $T\sim 100\,{\rm K}$, and ${\cal K}_e \sim 1$. This results in a relaxation time given by $1/T_{1,e}[{\rm s}^{-1}] \approx 10^3/R[{\rm nm}]^2$ and $T_{1,z}[{\rm s}^{-1}]\approx 5\times 10^4/R[{\rm nm}]^4$. We note that the relaxation times on the millisecond to $\sim 10$ second range can be accessed with current experimental setups using NV centers in diamond at temperatures around $100\,{\rm K}$.\cite{2012jarmola} We also note that, for typical probe-to-sample distances on the order of a few nanometers, charge and spin noise are comparable, making the spin degree of freedom of the probe essential to separate each contribution. Different noise components can be distinguished, for instance, by measuring relaxation time of differently oriented NV center probes in diamond.\cite{2010vectormagnetometer} 

\subsubsection{Noise as a function of distance}

For generic $R$ values, the behavior of magnetic noise as a function of distance is straight-forward to obtain because only a single $q$ wavevector is being sampled, $q = \omega/v_m$. For $R \lesssim v_m/\omega$, the form factor in the integral of Eq.(\ref{eq:noisell}) behaves as $[q^2F_e(q,0,R,\omega)]\approx 1/R^2$, such that the $R$ dependence can be factored out of the integral in Eq.(\ref{eq:noisell}), giving rise to a $1/R^2$ power-law of ${\cal N}_e$. Similar analysis is valid for spin noise which gives rise to a $1/R^4$ dependence. At intermediate to large distances, $R \gtrsim v_m/\omega$, noise in Eq.(\ref{eq:noisell}) is obtained by integrating  the product of the form factor $[q^2F_m(q,0,R,\omega)] \approx qe^{-2qR}/R$, valid for $qR \gtrsim 1$, and the spectral density which is a $\delta$-function at $q = \omega / v_m$. This yields charge and spin magnetic noise that falls off exponentially with a characteristic length $v_m / \omega$, ${\cal N}_m (R) \propto {\rm exp}({-2R\omega/v_m})/R$. 

\subsubsection{Noise as a function of temperature}

For clean systems, the temperature dependence of noise is governed by the ${\rm coth}(\hbar\omega/2k_{\rm B}T)$ factor in Eq.(\ref{eq:noisell}). In particular, for LLs, the spin probe samples ballistic charge and spin density waves with wavevector $q = \omega / v$. Within the bosonization description, such waves are non-interacting phonon-like modes with $T$-dependent amplitude $|\rho_{e,\sigma}| \propto \sqrt{T}$; this results in charge and spin-induced magnetic noise scaling as ${\cal N}_{e,\sigma}(T)\propto T$ in the semiclassical limit. The linear behavior with $T$ is valid up to $k_{\rm B} T \lesssim\hbar \omega$, where noise reaches a minimum value which is due to quantum fluctuations. The latter regime can be captured with NV centers, given that $\omega$ is on the range GHz-THz and can be comparable to $k_{\rm B}T$, and differs from NMR, which operates in the regime $\hbar\omega\ll k_{\rm B}T$. Both saturation and linear $T$ dependence is captured by the ${\rm coth}(\hbar\omega/2k_{\rm B}T)$ term. This simple temperature behavior is altered by disorder which introduces non-universal power-laws. 

\subsection{Case II: Helical and spin-polarized channels}

Spin-polarized states can occur in quantum wires in very strong magnetic fields ($\mu_{\rm B}B \sim E_{\rm F}$, with $E_{\rm F}$ the Fermi energy) or in the presence of ferromagnetic interactions with broken SU(2) symmetry---either due to easy axis or a magnetic field.\cite{2006ferromagneticedges,19961dspinpolarization} Helical states arise at the edges of 2D systems with strong spin-orbit coupling. Both spin-polarized and helical states introduce several qualitatively distinct behaviors, some of which are discussed in the present section and others in the context of disorder. The 1D motion of spin-polarized or helical fermions can be described with half as many degrees of freedom than in the SU(2) case. In either case, the 1D bosonic action is of the form
\be
{\cal S}_{\rm 1D} = \iint dt dx \left[\,i\Pi \partial_t \phi - {\cal H}_0(\phi,\Pi)\right].
\label{eq:action1dsp}
\ee
The bosonized degrees of freedom, $\Pi$ and $\phi$ are canonically conjugate, $[\phi(x),\Pi(x')] = i \delta(x-x')$ and describe charge/spin excitations of the 1D system. In the absence of disorder scattering, dynamics is governed by 
\be
{\cal H}_{0} (\phi,\Pi) = \frac{\hbar v_{\rm F}}{2\pi} \left[(\pi {\Pi})^2 + (\partial_x\phi)^2\right].
\label{eq:h1dspinlesssp}
\ee
Similar to the SU(2) case, here we do not include Coulomb repulsion because this is already accounted for in the full action, Eq.(\ref{eq:action}). 
In the long wavelength limit, charge density is related to the bosonized degrees of freedom via $\rho_e =-\partial_x\phi / \pi$. For the spin density $\rho_\sigma$, we assume for concreteness that the preferential direction is the $z$ axis (see Fig.\ref{fig:schematics}) such that $\rho_\sigma = \rho_z$. For spin-polarized states, charge and spin densities are equal, $\rho_e = \rho_\sigma = -\partial_x \phi/\pi$; for helical states, spin density is given by $\rho_\sigma = \Pi$. The reason behind this difference is more evident when carrier density is decomposed into its constituent flavors, namely fields $\phi_{r\sigma}$ corresponding to carriers with chirality $r = \pm$, spin polarization $\sigma = \uparrow,\downarrow$, and carrier density $\rho_{r\sigma} = r\partial_x \phi_{r\sigma}/\pi$. Within bosonization, it is standard to define $\phi = -(\phi_{+\sigma} - \phi_{-\sigma'})$ and $\Pi = \partial_x (\phi_{+\sigma} + \phi_{-\sigma'})/2\pi$. As such, for spin-polarized states, the total spin density is obtained by the sum in density of right-movers and left-movers, $\rho_\sigma = -\partial_x(\phi_{+\uparrow}+\phi_{-\uparrow})/\pi=\rho_e$. For helical states, on the other hand, the total spin density is obtained by the difference in density of right-movers and left-movers, $\rho_\sigma = \partial_x(\phi_{+\uparrow}-\phi_{-\downarrow})/2\pi=\Pi$. Excitations in $\rho_x$ and $\rho_y$ are assumed to be gapped and do not contribute to noise --- this is valid so long as $T$ is small compared to the energy scale of the polarizing mechanisms, e.g. ferromagnetism, spin-orbit coupling or Zeeman splitting. 

Integration of electromagnetic modes and calculation of charge and spin density correlations proceeds exactly as in the SU(2) case. For spin-polarized states, we find
\be
{\rm SP:}\quad{\cal C}_{\rho_m\rho_m}^{\rm R}(q,\omega) =\frac{1}{\pi}  \frac{{\cal K}_e v_e q^2 }{(v_e q)^2 -(\omega+i\epsilon)^2 },\quad m = e,\sigma,
\label{eq:rhorhosp}
\ee
whereas for helical states we find
\be
{\rm Helical:}\quad \begin{array}{c} \displaystyle {\cal C}_{\rho_e\rho_e}^{\rm R}(q,\omega) =\frac{1}{\pi}  \frac{{\cal K}_e v_e q^2 }{(v_e q)^2 -(\omega+i\epsilon)^2} \\ 
\displaystyle {\cal C}_{\rho_\sigma\rho_\sigma}^{\rm R}(q,\omega) =\frac{1}{\pi}  \frac{ v_e q^2/{\cal K}_e }{(v_e q)^2 -(\omega+i\epsilon)^2}
\end{array}.
\label{eq:rhorhoh}
\ee
In particular, charge correlations have the same form for the spin-polarized,  helical, and SU(2) states (albeit a factor of two due to spin degeneracy). Spin correlations, however, are different for each of these states. For spin polarized states, because $\rho_\sigma = -\partial_x\phi/\pi$, we find that the amplitude of spin fluctuations is proportional to ${\cal K}_e$. For helical states, because $\rho_\sigma = \Pi$, we find that the amplitude of spin fluctuations is proportional to $1/{\cal K}_e$. Analysis of the $R$ and $T$ dependence of noise from 
Eqs.(\ref{eq:rhorhosp}) and (\ref{eq:rhorhoh}) leads to the same conclusions as in the SU(2) case. The key distinction between SU(2), spin polarized, and helical states in the LL regime is how Coulomb repulsion affect spin noise, as discussed next. 

\subsubsection{Effect of repulsion strength on spin noise}
\label{sec:soc}

\begin{figure}
  \centering\includegraphics[scale=1.0]{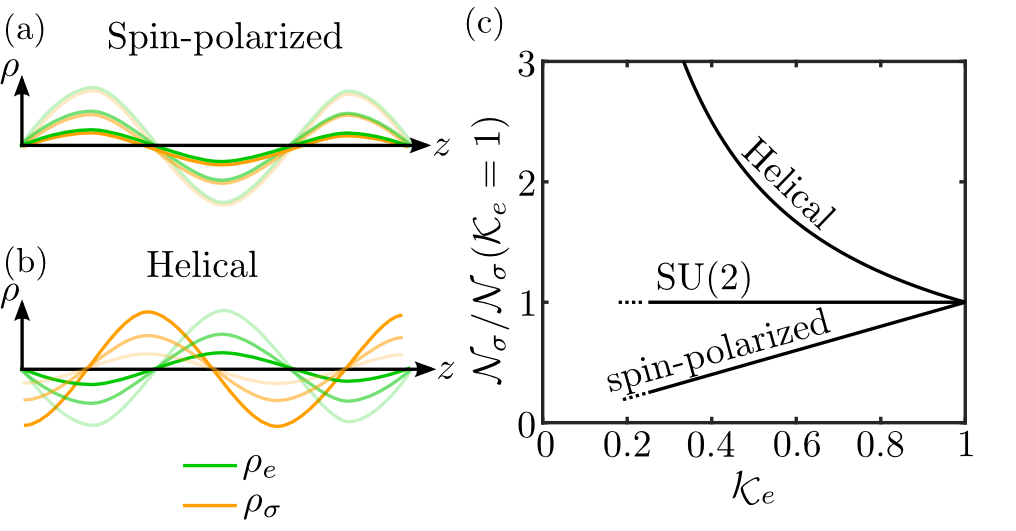}
  \caption{Effect of tuning repulsion strength on spin noise. (a) Spin fluctuations (orange) for spin-polarized states are locked to charge fluctuations (green); as such, both charge and spin fluctuations are suppressed with increasing Coulomb repulsion (increasing shading of lines). (b) For helical states, it is possible to have finite spin density in the absence of charge density; in this case, spin fluctuations are enhanced with increasing interaction strength. (c) By measuring spin noise while tuning the interaction strength it is possible to distinguish a helical, spin polarized or spinful metal. In the metallic case, because of charge-spin separation, spin noise is not sensitively affected by Coulomb repulsion.}
  \label{fig:interactionstrength}
\vspace{-4mm}
\end{figure} 

The internal structure of carriers has interesting manifestations in the spin noise behavior. We illustrate this effect by assuming in-situ control of ${\cal K}_e$ while measuring spin fluctuations. Control of the LL parameters ${\cal K}_e$ and $v_e$ can be achieved by using gate potentials.\cite{2005chargespinseparation} We recall from Eqs.(\ref{eq:rhorho}), (\ref{eq:rhorhosp}), and (\ref{eq:rhorhoh}) that the Luttinger parameter ${\cal K}_e$ affects the amplitude of spin fluctuations in different ways: spin fluctuations are proportional to ${\cal K}_e$ for spin-polarized states, $1/{\cal K}_e$ for helical states, and unity for SU(2) states. Physically, this characteristic spin noise behavior can be understood as follows: a spin density for spin-polarized states has to be accompanied by a charge density; as such, both charge and spin density fluctuations are suppressed for stronger Coulomb repulsion [Fig.\ref{fig:interactionstrength}(a)]. A spin density for helical states, on the contrary, can exist in the absence of a charge density; because spin and charge are conjugate fields, Coulomb suppression of charge fluctuations enhances spin fluctuations [Fig.\ref{fig:interactionstrength}(b)]. These two contrasting behaviors, furthermore, are distinct from that in a spinful 1D metal where spin noise is unaffected by long range Coulomb interactions. As a result, although fluctuations of the charged mode generically decrease at increasing values of ${\cal K}_e$, spin noise can either increase (helical phase), remain constant (spinful metal) or decrease (spin polarized phase) depending on the nature of 1D states (Fig.\ref{fig:interactionstrength}). 

\section{Noise from dirty wires: weak disorder}
\label{sec:noiseweakdisorder}

Scattering with a disorder potential couples right-movers to left-movers. There are two qualitatively distinct types of scattering behaviors which are usually encountered in 1D systems. The first case, which is discussed in the present section, is when impurities are dense and weak enough such that the effect of a single impurity is negligible but their collective effects are important (Gaussian disorder). The second case, discussed in the next section, is when impurities are scarce but strong (Poisson disorder). Regardless of the details of the scattering potential, it is known that disorder, no matter how weak, gives rise to strong deviations from LL behavior, e.g. Anderson insulators. 

Focusing on Gaussian disorder, it is known that charge density waves become pinned by the disorder potential below a characteristic pinning frequency $\omega_* $.\cite{1978fukuyamalee,19951dvariational} The value of $\omega_*$ is related to the localization length $\ell_{\rm loc}$ of electron wavefunctions via $\omega_* = v_e / \ell_{\rm loc}$. Within the long wavelength bosonization description, we can effectively incorporate the effects of disorder to describe quenching of long wavelength fluctuations due to pinning. This approach fails to describe physics occurring at lengthscales smaller than $\ell_{\rm loc}$, as will be described in more detail below. We begin by discussing noise in wires with an SU(2) degree of freedom and Gaussian disorder for $\omega\gtrsim\omega_*$, and compare the resulting noise behavior with that obtained for LLs in the previous section. Afterwards, we discuss qualitatively distinct behaviors that appear in 1D spin-polarized and helical states, also in the regime $\omega\gtrsim\omega_*$. In the final part of this section, we discuss qualitatively distinct behaviors that may arise in the pinned phase, $\omega \lesssim \omega_*$. 

\subsection{Case I: SU(2) channels with $\omega\gtrsim\omega_*$}

The starting point to discuss disordered wires is the LL Hamiltonian ${\cal H}_{m}$ [Eq.(\ref{eq:action1d})] describing ballistic propagation of charge and spin density waves. We introduce scattering via the disorder potential in bosonized form\cite{1988giamarchidisorder} 
\be
{\cal H}_{\rm dis}(\phi_e,\phi_\sigma) = \frac{u(x)}{\pi a} e^{i\sqrt{2} \phi_e(x)}\cos[\sqrt{2}\phi_\sigma(x)]+ {\rm h.c.},
\label{eq:disorder}
\ee
which describes spin-conserving backscattering. Here $a$ is the lattice cutoff and $u(x)$ is the continuum limit corresponding to the $2k_{\rm F}$ components of the scattering potential.\cite{giamarchibook,1988giamarchidisorder} The potential is assumed to be uncorrelated in space, $\langle {u}(x)u^*(x') \rangle = D \delta (x-x')$. Importantly, although the potential couples to $\rho_{e}$, the long wavevector backscattering components of the scattering potential also introduces scattering in the spin sector. 

The term ${\cal H}_{\rm dis}$ introduces competition between two opposite behaviors, namely ballistic propagation of density waves promoted by ${\cal H}_0$, and real-space pinning of the charge density promoted by ${\cal H}_{\rm dis}$. Pinning of charge density occurs because it is energetically favorable for $\phi_e$ to track the phase $\phi_{\rm imp}(x)$ of the impurity potential,  $u(x) = |u(x)|e^{i\phi_{\rm imp}(x)}$. Several approaches are available to describe scattering in an effective fashion both above and below $\omega_*$, each with its own limitations. To keep the formalism as generic as possible, here we use the memory function formalism, which has been commonly used to obtain correlations in a variety of non-Fermi liquids.\cite{1972goetzewolf,1988giamarchidisorder,1991giamarchiumklapp,2009violationwfl,2003reviewmemoryfunction} 

\subsubsection{Equations of motion}

To describe the effects of disorder, we derive the equations of motion within the memory function formalism, in order to give an intuitive picture of how microscopic dynamics is affected by disorder scattering. The key idea of this approach is to track only a subspace of relevant degrees of freedom, namely $[\phi_m(q,t),\Pi_m(q,t)]$ for a fixed value of $q$. The remaining degrees of freedom, which are orthogonal to the subspace spanned by $[\phi_m(q,t),\Pi_m(q,t)]$, enter the equations of motion via fluctuation and dissipation terms. It is clear from the Hamiltonian ${\cal H} = {\cal H}_0+{\cal H}_{\rm dis}$, that only ${\cal H}_{\rm dis}$ couples $[\phi_m(q,t),\Pi_m(q,t)]$ to their orthogonal subspace. In particular, the operator $f_m = [{\cal H}_{\rm dis},\Pi_m]$ quantifies the leakage out of the subspace spanned by $[\phi_m(q,t),\Pi_m(q,t)]$. Within this approach, the equations of motion $i\partial_t\Pi_m =\frac{1}{\hbar} [\Pi_m,{\cal H}]$ and $i\partial_t\phi_m =\frac{1}{\hbar} [\phi_m,{\cal H}]$ can be exactly expressed as:
\bea
\partial_t \phi_m (q,t) & = & \pi  {\cal K}_m v_m \Pi_m (q,t), 
\label{eq:eommm1}
\\ & & \nonumber\\ 
\partial_t \Pi_m(q,t) & = &  -(v_m q^2 / \pi{\cal K}_m)\phi_m (q,t)  \label{eq:eommm2}\\
 & & +  i \int_0^\infty ds{\cal M}_m(q,s) \Pi_m(q,t-s)  + \xi_m(q,t).\,\,\, \nonumber
\eea
For each value of $m = e,\sigma$, the equations of motion resemble those of a damped harmonic oscillator in which $\phi_m$ plays the role of position, $\Pi_m$ of momentum, and $1/\pi{\cal K}_mv_m$ of mass. Here ${\cal M}_m(q,t)={\cal M}_m'(q,t)+i{\cal M}_m''(q,t)$ is a memory function which introduces dissipation and retardation effects, and $\xi_m(q,t)$ is a random fluctuating force. We note that there is neither memory function nor force term in Eq.(\ref{eq:eommm1}) because the scattering potential depends on $\phi_m$, i.e. $ [{\cal H}_{\rm dis}(\phi_m),\phi_m] = 0 $, but not on $\Pi_m$. We also note that coupling between $\Pi_e$ and $\Pi_\sigma$ in Eq.(\ref{eq:eommm2}) is not present because $\langle f_e f_\sigma\rangle = 0$.

Although Eqs.(\ref{eq:eommm1}) and (\ref{eq:eommm2}) are exact,\cite{1965Mori} calculating the exact form of ${\cal M}_m$ and ${\xi}_m$ is challenging. For this reason, perturbation schemes have been developed to quantify such terms. \cite{1972goetzewolf} To leading order in $f_m$, ${\cal M}_m(q,\omega)$ is given by 
\be
{\cal M}_m(q,\omega) = \frac{{\cal C}_{f_mf_m}^{\rm R}(q,\omega) - {\cal C}_{f_mf_m}^{\rm R}(q,0)}{\omega}.
\label{eq:memoryfunctiondef}
\ee
In thermal equilibrium, the fluctuating force and the memory function are not independent. In particular, fluctuations are related to dissipation via the fluctuation-dissipation theorem: $\int_{-\infty}^{\infty}dt \langle \xi_m(-q,t)\xi_m(q,0)\rangle e^{i\omega t}= 2k_{\rm B}T{\cal M}_m''(q,\omega) / \pi \hbar v_m {\cal K}_m$.

Interpretation of Eqs.(\ref{eq:eommm1})-(\ref{eq:eommm2}) is straightforward when they are rewritten in terms of charge and currents in order to obtain hydrodynamic equations. In particular, by multipying Eq.(\ref{eq:eommm1}) by $\sqrt{2}q/\pi$, using $\rho_m = -\sqrt{2}\partial\phi_m / \pi$ and $j_m = \sqrt{2}v_{\rm F}\Pi_m = \sqrt{2}{\cal K}_mv_m \Pi_m$, we recover the continuity equation, $\partial_t\rho_m = v_m\partial_x j_m$ for charge and spin modes. Similarly, by multiplying Eq.(\ref{eq:eommm2}) by $v_m$ and assuming for illustrative purposes that the memory function is local in time and purely imaginary, i.e. $i \int_0^\infty ds{\cal M}_m(q,s) \Pi_m(q,t-s) = -\nu j_m(q,t)$, we obtain an equation describing current dynamics in a resistive circuit: $\partial_t j_m = v_m^2\partial_x\rho_m -\nu j_m(x,t) +\xi_m(t) $: the first term in the right hand side is the driving force for the current (note that $\hbar v_e\partial_x\rho_m = \partial_x \mu$ is the gradient of the chemical potential), the second term is a resistive/dissipative term, and the third term is a random, fluctuating force. 

From the equations of motion (\ref{eq:eommm1}) and (\ref{eq:eommm2}), it is possible to compute ${\cal C}_{\phi_m\phi_m}^{\rm R}(q,\omega)$, from which it is then trivial to obtain ${\cal C}_{\rho_m\rho_m}^{\rm R}(q,\omega)$. The first step in this direction is to go into Fourier space and invert Eqs.(\ref{eq:eommm1}) and (\ref{eq:eommm2}):
\be
\left(\begin{array}{c} \phi_m \\\Pi_m \end{array}\right) = \frac{\xi_m(q,\omega)}{(v_mq)^2 - \omega^2 - \omega{\cal M}_m (q,\omega)}\left(\begin{array}{c} \pi{\cal K}_m v_m\\ -i\omega \end{array}\right).
\label{eq:eom3}
\ee
Obtaining ${\cal C}_{\phi_m\phi_m}^{\rm R}(q,\omega)$ via Eq.(\ref{eq:eom3}) proceeds in two steps. 
First, we express the product $\langle\phi_m(q,\omega)\bar{\phi}_m(q,\omega)\rangle$ in terms of $\langle \xi_m(q,\omega)\bar{\xi}_m(q,\omega)\rangle$, which can be obtained from the fluctuation-dissipation theorem: $\langle \xi_m(q,\omega)\bar{\xi}_m(q,\omega)\rangle = 2k_{\rm B}T{\cal M}_m''(q,\omega) / \pi \hbar v_m {\cal K}_m$. Second, we make the connection between correlation functions $\langle\phi_m(q,\omega)\bar\phi_m(q,\omega)\rangle = \frac{2k_{\rm B}T}{\hbar \omega}{\rm Im}[{\cal C}_{\phi_m\phi_m}^{\rm R}(q,\omega)]$, which is valid in thermal equilibrium and in the classical limit $k_{\rm B}T \gg \hbar\omega$. These two steps result in
\be
{{\cal C}}_{\phi_m\phi_m}^{\rm R}(q,\omega) = \frac{\pi {\cal K}_m v_m}{(v_mq)^2 - \omega^2 - \omega{\cal M}_m(q,\omega)}.
\label{eq:phiphi}
\ee
From here, it is trivial to obtain charge and spin density correlators by using the mapping $\rho_m = \sqrt{2}\partial_x\phi_m/\pi$, which results in 
\be
{\rm SU(2)}:{\cal C}_{\rho_m\rho_m}^{\rm R}(q,\omega) = \frac{2}{\pi}\frac{{\cal K}_m v_m q^2}{(v_mq)^2 - \omega^2 - \omega{\cal M}_m(q,\omega)}. 
\label{eq:rhorhodisordersf}
\ee 

\subsubsection{Memory functions for disorder scattering}

Because the disorder potential is short-ranged and Gaussian, ${\cal M}_m(q,\omega)$ does not depend explicitly on $q$ (see Appendix\,\ref{app:memoryfunction}); ${\cal M}(q,\omega)$ does have, however, an implicit $q$-dependence via ${\cal K}_e$ and $v_e$ [c.f., Eq.(\ref{eq:luttingerparameters})]. Further, because ${\cal M}_m(q,\omega)$ depends on the temperature of the system, hereafter we show these dependencies explicitly, ${\cal M}_m(q,\omega) \equiv{\cal M}_m(q,\omega,T)$. 

The memory function can be generically written as 
\be
{\cal M}_m (q,\omega,T) = {\Gamma}_m \left( \frac{a k_{\rm B}T}{\hbar v_m}\right)^{\alpha_m}{\cal F}_m\left( \frac{\hbar \omega}{k_{\rm B}T}\right),\,m=e,\sigma,
\label{eq:memoryfunction}
\ee 
where $\Gamma_m$ is a constant with units of ${\rm sec}^{-1}$, $\alpha_m$ is a number that depends on microscopics, and ${\cal F}_m(x)$ is a dimensionless complex function. The details of the memory functions calculation are given in Appendix\,\ref{app:memoryfunction}. Given that spin probes usually operate with $\omega$ below THz frequencies ($\lesssim 4\,{\rm meV}$) and temperatures can vary over a wide range including room temperatures, we focus on the regime $\hbar\omega \lesssim k_{\rm B}T$ which is mostly relevant to experiments. In this regime, we find that ${\cal F}_m(x\lesssim 1)$ is approximately constant and, therefore, the sign of $\alpha_m$ determines whether the scattering rate increases or decreases with temperature. In particular, the memory function behaves as ${\cal M}_m(q,\omega,T) \approx i \beta \Gamma_m (a k_{\rm B}T/\hbar v_m)^{{\cal K}_e + {\cal K}_\sigma - 2}$, where $\Gamma_e = (2\pi)^{{\cal K}_e + {\cal K}_\sigma +1}D{\cal K}_e^2(v_e/v_\sigma)^{{\cal K}_\sigma}$ and $\Gamma_\sigma = (2\pi)^{{\cal K}_e + {\cal K}_\sigma +1}D{\cal K}_\sigma^2(v_\sigma/v_e)^{{\cal K}_e}$ (the values of ${\cal K}_e$ and $v_e$ are $q$-dependent). Because for repulsive interactions ${\cal K}_e + {\cal K}_\sigma -  2 <0 $ is valid, ${\cal M}_m(q,T)$ (and the scattering rate) monotonically decreases as a function of $T$. As such, the system behaves more `ballistic'-like as temperature increases. 

\subsubsection{Magnetic noise from disordered wires}

Combining Eqs.(\ref{eq:rhorhodisordersf}) and (\ref{eq:noise2}), we find charge and spin noise in a disordered wire given by 
\be
\begin{array}{l}
\displaystyle{\cal N}_m (\omega,T,R)= \displaystyle\frac{(g_{\rm s}\mu_{\rm B})^2}{8\pi}\int_0^\infty \frac{dq}{2\pi} F_m(q,0,R,\omega) \\  \displaystyle\quad\quad \frac{ {\rm coth}(\hbar\omega / 2k_{\rm B} T) {\cal K}_m v_m q^2 \omega{\cal M}_m''(q,\omega,T)}{[(v_mq)^2 - \omega^2 -\omega{\cal M}'(q,\omega,T) ]^2 + [\omega{\cal M}_m''(q,\omega,T)]^2}.
\end{array}
\label{eq:noisedisspinful}
\ee
Equation (\ref{eq:noisedisspinful}) summarizes the essence of magnetic noise measurements in the presence of disorder. In particular, disorder couples modes with different $q$ wavevectors and, as such, charge and spin fluctuations at frequency $\omega$ are distributed in $q$-space. This is qualitatively distinct from clean systems where fluctuations of frequency $\omega$ are dominated by a single wavevector $\omega/v_m$. 

In order to simplify the discussion, in what follows we fix the values of ${\cal K}_e$ and $v_e$ appearing in ${\cal M}_m$ such that ${\cal M}_m$ no longer depends implicitly on $q$. This approximation is exact when scattering rate is small, $\omega \gtrsim {\cal M}_m''$, given that most of the contribution to noise in Eq.(\ref{eq:noisedisspinful}) comes from a small phase space region centered at $q = \omega / v_e$ for ${\cal N}_e$, and $q = \omega / v_{\rm F}$ for ${\cal N}_\sigma$. For large scattering rate, $\omega\lesssim {\cal M}''$, a larger region in $q$ space contributes to noise and, as such, there will be $q$-dependent logarithmic corrections to ${\cal K}_e$ and $v_e$. We neglect these secondary corrections, which do not alter the qualitative behavior, but that could easily be incorporated if a detailed quantitative analysis were needed. The distribution of modes in $q$-space modifies the dependence of magnetic noise as a function of distance found for LLs and, furthermore, results in non-universal power-law of $T$, as will be discussed next. 

\subsubsection{Noise as a function of distance}
\label{sec:distance}

\begin{figure}
  \centering\includegraphics[scale=1.0]{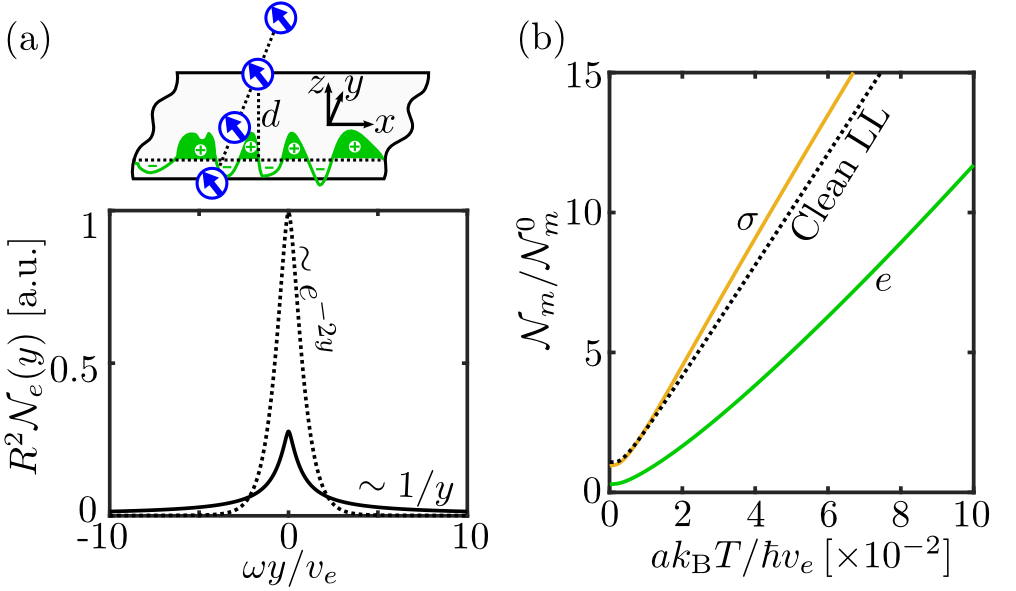}
  \caption{(a) Qualitatively distinct noise behaviors as a function of probe-to-sample distance for clean (dotted line) and disordered (solid line) systems. Clean systems exhibit an exponentially decay of noise as a function of $R$ governed by the lengthscale $v_e/\omega$. For disordered systems, charge fluctuations of frequency $\omega$ are comprised of modes several modes distributed in $q$-space. In this case, we find noise decaying as $y^{-3}$ power law (see discussion in main text). Here we assume that the spin probe is positioned at ${\bm r} = (0,y,d)$ and the 1D system runs in the ${\bm r}=(x,0,0)$ line. (b) The temperature dependence of charge ($m=e$) and spin ($m=\sigma$) noise exhibits non-universal power-law behavior. Backscattering leads to localization and suppression of charge and spin density fluctuations at $T=0$. Temperature combined with interactions makes the system less susceptible to scattering and enhances magnetic noise. For $\hbar\omega\lesssim k_{\rm B}T $, we find superlinear(sublinear) $T$-dependence of charge(spin) noise. Tuning sample-to-probe distance as a diagnostics of transport regime. 
}
  \label{fig:noise-vs-distance}
\vspace{-4mm}
\end{figure} 

We begin the analysis of Eq.(\ref{eq:noisedisspinful}) by exploring how modes in disordered wires are sampled as the probe-to-sample distance is changed. Let us focus on charge-induced noise, and focus on the regime $k_{\rm B}T/\hbar \gtrsim \omega \gtrsim \omega_*$, where it is valid to assume that ${\cal M}_m''(\omega,T) \gg {\cal M}_m'(\omega,T)$ [see paragraph following Eq.(\ref{eq:memoryfunction})]. We also assume that scattering rate is large, $\omega\lesssim{\cal M}_m''(\omega,T)$, such that density waves start to become pinned by disorder and depart from the LL behavior discussed in the previous section. Figure\,\ref{fig:noise-vs-distance}(a) shows charge-induced noise as a function of distance from the disordered wire, and the results are compared to those for clean wires. We recall that, for LLs, noise decays as $1/R^2$ until $R$ becomes comparable to the excitations wavelength $\omega/v_e$; beyond this length, noise decays exponentially as a function of distance. For disordered wires at close proximity we find that magnetic noise also decays as $1/R^2$ so long as $R\lesssim \ell_{\rm d} = v_m/\sqrt{\omega{\cal M}_m}$, where $\ell_{\rm d}$ is the scattering length, but its overall magnitude is smaller due to charge pinning. Interestingly, we find that for $R\gtrsim \ell_{\rm d}$, magnetic noise in a disordered wire decays as a $1/R^3$ power law and can overwhelm magnetic noise for clean wires, which decays exponentially with $R$. 

This behavior can be understood as follows. In the regime in which backscattering dominates, the denominator of the integrand in Eq.(\ref{eq:noisedisspinful}) is dominated by the scattering rate $\omega{\cal M}_m''$, and can be approximated as $\sim 1/[(v_mq)^4+(\omega{\cal M}_m'')^2]$. As such, the noise integral in Eq.(\ref{eq:noisedisspinful}) can be written as $\int_0^\infty dq F_e(q,0,R,\omega) {\cal M}_m''/[(v_mq)^4+(\omega{\cal M}_m'')^2]$. For $R \lesssim \ell_{\rm d}$, we can use $F_e(q,0,R,\omega) \sim {1}/(qR)^2$, from which a $1/R^2$ dependence is obtained by pulling the $1/R^2$ factor out of the integral. For $R \gtrsim \ell_{\rm d}$, we can make the integral dimensionless by defining $x = qR$ and using $\omega{\cal M}_e''/R^2 \ll 1$ in the denomiator, which results in a $1/R^3$ power-law behavior, ${\cal N}_e = [e^2 {\cal K}\omega^2 \ell_{\rm d}^2/ 8\pi^2v^2R^3] \int_0^\infty dx x^2K_1^2(x)$ [see Fig.\ref{fig:noise-vs-distance}(b)]. As such, the exponential vs. power-law behavior of noise as a function of probe-to-sample distance can be used as diagnostics of pinning of density waves. 

\subsubsection{Noise as a function of temperature: emergence of non-universal power laws}
\label{sec:disordered}

Disorder results in the emergence of non-universal power-laws of $T$, one of the key signature of LL. Again, let us focus on the regime $k_{\rm B}T/\hbar \gtrsim \omega \gtrsim \omega_*$, where ${\cal M}_m(\omega,T)$ is strongly $T$-dependent and it is valid to assume ${\cal M}_m''(\omega,T) \gg {\cal M}_m'(\omega,T)$, and that the scattering rate is large, ${\cal M}_m''(\omega,T)\gtrsim\omega$, such that we are away from LL behavior. Figure \ref{fig:noise-vs-distance}(b) shows the dependence of charge and spin noise as a function of $T$, and the results are compared with those of clean LLs which is governed by a ${\rm coth}(\hbar\omega/2k_{\rm B}T)$ factor. Results are plotted assuming that the probe is close to the sample, $R/\ell_{\rm dis} \lesssim 1$. We find that charge and spin noise have non-universal power law behavior, with charge noise behaving superlinerly, whereas spin noise behaves sublinearly. This behavior can be understood as follows. In the regime in which backscattering dominates, the denominator in Eq.(\ref{eq:noisedisspinful}) can be approximated using  $1/[(v_mq)^4+(\omega{\cal M}_m'')^2]$ and the integral in $q$-space can be expressed as ${\cal N}_m \propto \int_0^\infty dq q^r/[(v_mq)^4 + (\omega{\cal M}_m'')^2]$, where $r=0$ for $m=e$ and $r=2$ for $m=\sigma$. The noise integral can be made dimensionless by defining  $x = v_m q / \sqrt{\omega{\cal M}_m''} $ such that all $T$ dependent terms appear as prefactors of the integral. Upon normalization, Eq.(\ref{eq:noisedisspinful}) gives rise to noise scaling with temperature as ${\cal N}_e(T) \propto  T / \sqrt{{\cal M}_m''(T)} $, and spin noise scaling as ${\cal N}_\sigma(T) \propto T \sqrt{{\cal M}_m''(T)}$, where the factor $T$ is introduced by the ${\rm coth}(\hbar\omega/2k_{\rm B}T)\approx 2k_{\rm B}T / \hbar \omega$ term in the numerator of Eq.(\ref{eq:noisedisspinful}). Using ${\cal M}_m(\omega,T)$ in Eq.(\ref{eq:memoryfunction}) for $k_{\rm B}{T}\gtrsim \hbar\omega$, we find that, when scattering rate is large, charge-induced noise increases superlinearly with $T$, ${\cal N}_e \propto T^{2-({\cal K}_e + {\cal K}_\sigma)/2}$, whereas spin noise increases sublinearly with $T$, ${\cal N}_\sigma \propto T^{({\cal K}_e +{\cal K}_\sigma)/2}$. We also note that charge fluctuations depend on ${\cal K}_\sigma$ (and viceversa) because the scattering potential in Eq.(\ref{eq:disorder}) couples the charge and spin degrees of freedom. 

\subsection{Case II: Helical and spin-polarized channels with $\omega\gtrsim\omega_*$}

The starting point to discuss disordered spin-polarized and helical states is the LL Hamiltonian, Eq.(\ref{eq:h1dspinlesssp}), describing ballistic propagation of charge/spin density waves. Proceeding with our minimal approach, we introduce scattering via a disorder potential of the form\cite{2006qshmoore} 
\be
{\cal H}_{\rm dis}(\phi) = \frac{u_n(x)}{\pi a} e^{2 i n \phi(x)}+ {\rm h.c.}
\label{eq:disordersp}
\ee
Here $a$ is the lattice cutoff and $u_n(x)$ is an uncorrelated potential $\langle {u}_n(x)\bar{u}_n(x') \rangle = D_n \delta (x-x')$. The value of $n=1,2$ reflects the amount of particles involved in scattering and captures two qualitatively distinct noise behaviors. The case $n=1$ corresponds to the usual direct backscattering term where $u_1(x)$ is the continuum limit of the $2k_{\rm F}$ components of the scattering potential.\cite{giamarchibook} It is often the case, however, that symmetries of the Hamiltonian do not allow such terms, e.g. in the quantum spin Hall states wherein helical states are protected from backscattering by time-reversal symmetry. Rather than specifying one of the several microscopic models which have been proposed,\cite{2009qshkondo,2012qshbackscattering,2013helicalresistance} here instead we capture scattering phenomenologically by using $n=2$ in Eq.(\ref{eq:action1d}), resembling two particles participating in the backscattering process, and $u_2(x)$ is an effective potential induced by second order processes. As we will see, the key effect of $n$ is to describe whether temperature combined with interactions enhances or quenches backscattering. 

The equations of motion for the helical and spin-polarized states are the same as those for the SU(2), Eqs.(\ref{eq:eommm1}) and (\ref{eq:eommm2}) but, because the spin degree of freedom is frozen, restricted to the charge sector $m=e$ . As such, we do not repeat the same procedure that lead to Eqs.(\ref{eq:eommm1})-(\ref{eq:rhorhodisordersf}), but only quote the final result. For spin-polarized states, we find 
\be
{\rm SP}: \quad {\cal C}_{\rho_m\rho_m}^{\rm R}(q,\omega,T) = \frac{1}{\pi}\frac{{\cal K}_e v_e q^2}{(v_eq)^2 - \omega^2 - \omega{\cal M}_1(\omega,T)},
\label{eq:rhorhodisordersp}
\ee
where we used the relation $\rho_e = \rho_\sigma$, and ${\cal M}_1$ is the memory function corresponding to the scattering potential in Eq.(\ref{eq:disordersp}) for $n=1$. For helical states we have to keep in mind that the correlator ${\cal C}_{\rho_\sigma\rho_\sigma}^{\rm R}(q,\omega)$ is obtained from ${\cal C}_{\Pi\Pi}^{\rm R}(q,\omega)$, see Eq.(\ref{eq:eom3}), which results in 
\be
{\rm Helical}: \,\,\begin{array}{l}
\displaystyle{\cal C}_{\rho_e\rho_e}^{\rm R}(q,\omega,T) = \frac{1}{\pi}\frac{{\cal K}_e v_e q^2}{(v_eq)^2 - \omega^2 - \omega{\cal M}_2(\omega,T)},\\
\displaystyle{\cal C}_{\rho_\sigma\rho_\sigma}^{\rm R}(q,\omega,T) =\frac{1}{\pi} \frac{\omega^2 / {\cal K}_e v_e}{(v_eq)^2 - \omega^2 - \omega{\cal M}_2(\omega,T)}.
\end{array}
\label{eq:rhorhodisorderh}
\ee
Here ${\cal M}_2(q,\omega)$ is the memory function corresponding to Eq.(\ref{eq:disordersp}) for $n=2$. 

To give a qualitative picture of the temperature dependence of ${\cal M}_n(\omega,T)$, we quote the results in the regime $\hbar\omega \lesssim k_{\rm B}T$ (details for all values of $\omega$, $T$ are described in the Appendix\,\ref{app:memoryfunction}). In this regime, we find ${\cal M}_n(\omega,T) \approx i \gamma_{n} (a k_{\rm B}T/\hbar v_e)^{2n{\cal K}_e-2}$, where $\gamma_n = (2\pi)^{2n{\cal K}_e -2}(D_n{\cal K}_e/v_e)$. For $n=1$, any value of repulsive interaction makes the scattering rate monotonically decreasing as a function of $T$; this behavior is the same as in the SU(2) case. This indicates that interactions combined with temperature tend to make the system less sensitive to the disorder. For $n=2$, there is a transition in the temperature-dependence of the ${\cal M}_2(\omega,T)$ which ocurrs at ${\cal K}_{\rm c} = 1/2$: scattering is enhanced (suppressed) at larger temperatures for ${\cal K}_e > {\cal K}_{\rm c}$(${\cal K}_e < {\cal K}_{\rm c}$). The existence of a critical repulsion strength which changes the importance of scattering at small temperatures is consistent with proposed microscopic models of scattering in quantum spin Hall phases (the value of ${\cal K}_{\rm c}$, however, is model specific). 

The analysis of how noise varies as a function of distance for disordered wires leads to the same power-law behaviors as those in the SU(2) case described in the previous section, so we do not reproduce the results here. Instead, here we focus on how single particle backscattering and interaction-assisted backscattering lead to qualitatively distinct noise behaviors as a function of $T$. 

\subsubsection{Noise as a function of temperature} 

\begin{figure}
  \centering\includegraphics[scale=1.0]{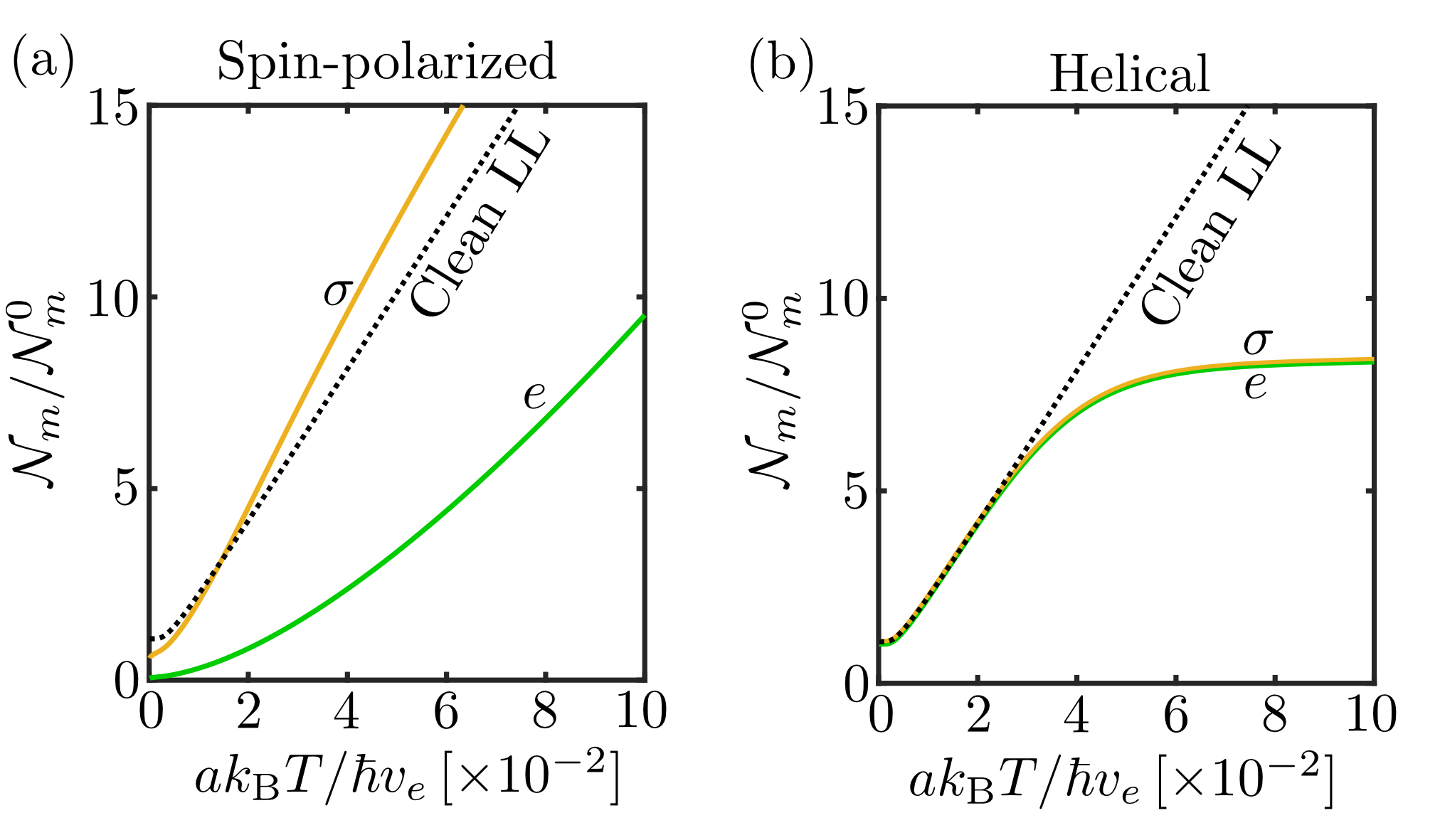}
  \caption{The temperature dependence of charge ($m=e$) and spin ($m=\sigma$) noise exhibits non-universal and qualitatively distinct behaviors for different microscopic phases and scattering mechanisms. (a) For the disordered spin-polarized phase at $T=0$, backscattering leads to localization and suppression of charge and spin density fluctuations. Temperature combined with interactions makes the system less susceptible to scattering and enhances magnetic noise. For $\hbar\omega\lesssim k_{\rm B}T $, we find superlinear(sublinear) $T$-dependence of charge(spin) noise. (b) In the helical phase, because of time-reversal symmetry, disorder backscattering is suppressed at $T=0$ and the system behaves as a perfect LL. Temperature combined with interactions assists disorder backscattering. For weak repulsion, we find sublinear(superlinear) $T$-dependence of charge(spin) noise, see details in main text. In the figures we use $ a \omega /v = 5\times10^{-3}$.
}
  \label{fig:disorder}
\vspace{-4mm}
\end{figure} 

We consider first the case of single particle backscattering, $n=1$ in Eq.(\ref{eq:disorder}), wherein temperature makes modes less sensitive to disorder. As before, we focus on the regime $k_{\rm B}T/\hbar \gtrsim \omega \gtrsim \omega_*$, where ${\cal M}_1(\omega,T)$ is strongly $T$-dependent and it is legitimate to assume ${\cal M}_1''(\omega,T) \gg {\cal M}_1'(\omega,T)$ [see paragraph following Eq.(\ref{eq:memoryfunction})], and also that scattering rate is large, $\omega\lesssim{\cal M}_1''(q,\omega)$, such that we deviate from LL behavior discussed in the previous section. Figure \ref{fig:disorder}(a) shows the dependence of charge and spin noise as a function of $T$ for the spin-polarized case, and the results are compared with those of clean LLs. The $T$ dependence of noise follows the same behavior as in the disordered wire with an SU(2) degree of freedom, see Fig.\ref{fig:noise-vs-distance}(b), but with a power law that depends only on ${\cal K}_e$. In particular, upon normalization and pulling out all the $T$-dependent terms of the integral, Eq.(\ref{eq:noisell}) gives rise to noise scaling with temperature as ${\cal N}_e(T) \propto  T / \sqrt{{\cal M}_1''(T)} $, and spin noise scaling as ${\cal N}_\sigma(T) \propto T \sqrt{{\cal M}_1''(T)}$. Using ${\cal M}_1(\omega,T)$ described in the previous section for $k_{\rm B}{T}\gtrsim \hbar\omega$, we find that charge-induced noise increases superlinearly with $T$, ${\cal N}_e \propto T^{2-{\cal K}_e}$, whereas spin noise increases sublinearly with $T$, ${\cal N}_\sigma \propto T^{{\cal K}_e}$ [see Fig.\ref{fig:disorder}(a) at moderate to high $T$]. 

Helical states exhibit qualitatively distinct noise behavior as a function of $T$ than the spin-polarized case for weak repulsion, ${\cal K}_e > 1/2$, see Fig.\ref{fig:disorder}(b). For small $T$, time-reversal symmetry protects chiral states against backscattering; this leads to a clean LL behavior at $T=0$. At large $T$, multiparticle interactions {\it assist} disorder in backscattering chiral states, resulting in an enhancement of the scattering rate. To roughly estimate the $T$ dependence when scattering rate is strong, we note the magnetic noise integral in $q$-space in Eq.(\ref{eq:noisedisspinful}) can be approximated as $\int_0^\infty dq {\cal M}_2''/[(v_mq)^4+(\omega{\cal M}_2'')^2]$, where we used the same approximations as in the previous paragraph. The key distinction with the spin-polarized case is that ${\cal C}_{\rho_\sigma\rho_\sigma}^{\rm R}$ in Eq.(\ref{eq:rhorhodisordersp}) contains an $\omega^2$ term in the numerator rather than a $q^2$ term, thus the sampling weight in $q$-space is different than in the spin-polarized case. As such, we find charge and spin noise scaling with temperature as ${\cal N}_{e,\sigma}(T) \propto  T / \sqrt{{\cal M}_2''(T)} $. For $\hbar\omega \lesssim k_{\rm B}T$, we find a sublinear $T$-dependent behavior for charge and spin noise, ${\cal N}_{e,\sigma}(T)\propto T^{2(1-{\cal K}_e)}$. 

A subtle yet interesting effect is that, because spin fluctuations are locked to charge fluctuations for spin-polarized and helical states, charge and spin noise as a function of temperature are also locked. Using the scaling as a function of $T$ found above, for spin polarized states, we find that the product ${\cal N}_{e}(T) {\cal N}_\sigma(T) / T^2 $ is independent of $T$, whereas for helical states, the product ${\cal N}_e (T)/ {\cal N}_\sigma(T)  $ is independent of $T$. This behavior suggests a diagnostics of helical {\it vs}. spin-polarized 1D channels. A similar behavior was discussed above for the SU(2) case for strong disorder, where the charge and spin sectors become coupled by the scattering potential, Eq.(\ref{eq:disordersp}). For more general 1D systems where charge and spin degree of freedom are separated and subject to different scattering potentials, e.g. Hubbard models, charge and spin noise are no longer locked and neither ${\cal N}_e(T) {\cal N}_\sigma(T) / T^2 $ nor ${\cal N}_e (T)/ {\cal N}_\sigma(T)  $ are independent of $T$. 

\subsection{Case III: 1D channels with $\omega\lesssim\omega_*$}

While memory functions correctly capture qualitatively behaviors which are important for our discussion, such as enhancement/quenching of scattering as a function of temperatures, there are more accurate approaches to describe the dependence on $T$ and $\omega$ particularly in the regime $\omega \lesssim \omega_*$. The key limitation of Eq.(\ref{eq:memoryfunctiondef}) is that, as soon as $\omega$ approaches $\omega_*$, higher order corrections (in powers of $f_m$) become necessary. One way to tackle this problem is to combine the memory function formalism with RG,\cite{1988giamarchidisorder} such that the microscopic parameters ${\cal K}_m$, $v_m$, and $D_m$, which are constant in our model, become $T$-dependent. This approach, however, fails to describe physics on the scale of $\ell_{\rm loc}$ which are important if the probe is located within $R \lesssim \ell_{\rm loc}$.

Another approach, which is valid when $k_{\rm B}T\lesssim\hbar\omega\ll\omega_*$, is the Gaussian variational approach, which consists of finding the best quadratic approximation to the disordered Hamiltonian via minimization in replica-space, in order to compute two-point correlations.\cite{19951dvariational} Replica-symmetry-breaking generates a mass term in the excitation spectrum, which can be described by replacing $-\omega{\cal M} \rightarrow M + i\gamma\omega$ in Eq.(\ref{eq:rhorhodisordersp}), where $M$ is a mass term and $\gamma$ is a $T$-dependent factor. The imaginary term $i\gamma\omega$ gives rise to the characteristic $\sigma(\omega) \propto \omega^2$ which governs conductivity in 1D systems, and qualitatively agrees with the usual ${\sigma}(\omega)\sim \omega^2{\rm ln}^2\omega$ obtained in the Anderson insulating regime (${\cal K}_e = 1$) as well as in the Fukuyama-Lee regime (${\cal K}_e = 0$).\cite{1978fukuyamalee} Contrary to RG, this approach fails to account for the renormalization of the Luttinger parameters. Further, it also fails to describe charge motion via quantum creep, which gives rise to variable range hopping.\cite{2003quantumcreepgiamarchi,2006quantumcreepgiamarchi}

\begin{figure}
  \centering\includegraphics[scale=1.0]{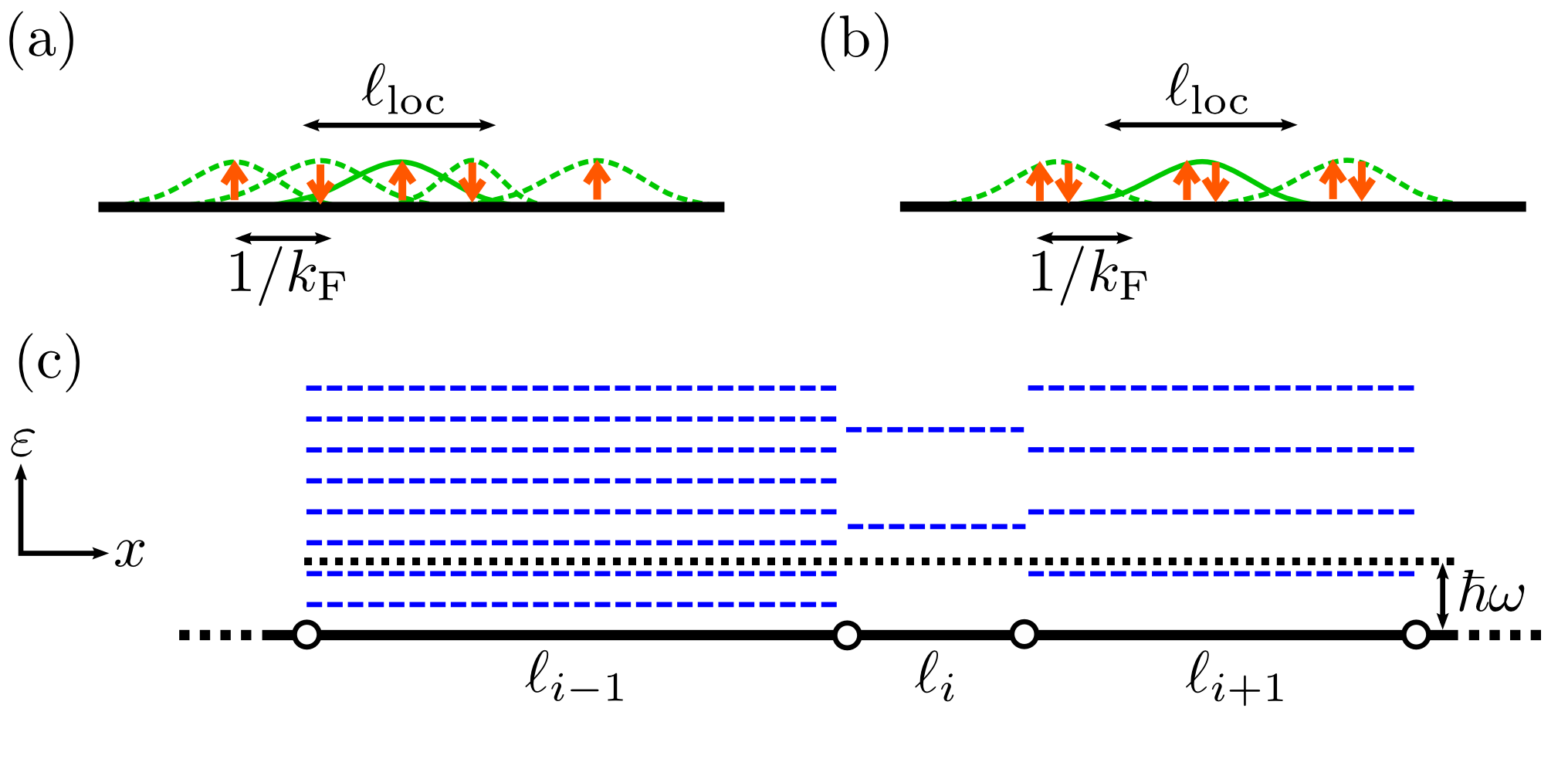}
  \caption{(a) Pinning of wavefunctions leads to quenching of charge fluctuations. For strong repulsion, neighboring spins are coupled via an antiferromagnetic coupling and spin fluctuations are possible. (b) For weak repulsion, spin form singlet states and spin fluctuations become gapped. (c) Spatial inhomogeneity and detuning of noise in the strong impurity regime. For strong, sparse impurities (empty circles), the wire is cut into segments of size $\ell_i$ and finite-size quantization effects, $\Delta\varepsilon_i = \hbar v / \ell_i$, take place. If the level-splitting of the probe $\omega$ is sufficiently detuned from $\Delta\varepsilon_i$, noise becomes negligibly small.
}
  \label{fig:strongimpurity}
\vspace{-4mm}
\end{figure}

Both RG and the Gaussian variational approach are good to describe quenching of long wavelength charge and spin fluctuations. Both approaches, however, rely on self-averaging of the disorder potential. This approximation may become questionable when accessing dynamics at distances $R \lesssim \ell_{\rm loc}$. In particular, while we still expect orbital degrees of freedom to be become frozen as the wavefunction becomes localized in space due to disorder, the spin degree of freedom can still fluctuate; long wavelength spin fluctuations are quenched, but spin fluctuations on the lattice scale remain, e.g. paramagnetic fluctuations. If Coulomb repulsion is sufficiently strong, interactions between neighboring spins are antiferromagnetic, and each spin interacts with $\sim k_{\rm F}\ell_{\rm loc}$ nearest neighbors via random exchange parameters, see Fig.\ref{fig:strongimpurity}(a). This creates islands -- or rare regions -- where spins are strongly correlated (for a general discussion of spin fluctuations in disordered systems, see Ref.[\onlinecite{1992spindisorder}]). As such, spin probes can detect spin-induced noise induced by paramagentic fluctuations in these islands, so long as the probe-to-sample distance is on the order of island size. For weak repulsion, it has been shown that localization can also lead to localized singlet states in which spin fluctuations become gapped, see Fig.\ref{fig:strongimpurity}(b).\cite{1988giamarchidisorder} In this case, spin noise is quenched even on the scale $\ell_{\rm loc}$.

\section{Noise from dirty wires: strong impurties}
\label{sec:noisestrongdisorder}

A different disorder regime is present when impurities introduce strong, local scattering potentials which are, on average, sufficiently separated from each other. In this regime, a scattering potential of the form $U(x) = \sum_i u_i \delta(x-x_i)$ is assumed, such that the separation of impurities is large on the lattice and excitation wavelength scale, $\ell_i = x_{i+1} - x_{i} \gtrsim v/\omega \gg a$. Further, we assume that $u_i/ a$ is weak compared with Fermi energy of 1D electronic states such that bosonization is still valid, but large compared to $\omega$. Let us focus first on the spin-polarized phase. Within the bosonization description, the impurity Hamiltonian introduced by $U(x)$ can be written as 
\be
{\cal H}_{\rm imp} = \sum_i \frac{u_i}{\pi a} \cos[2\phi(x = x_i)], 
\label{eq:kanefisher}
\ee
where unimportant forward-scattering terms are removed. Following Kane and Fisher,\cite{1992kanefisher1,1992kanefisher2}
the key effect of Eq.(\ref{eq:kanefisher}) is to pin $\phi(x)$ to the impurity potential at the positions $x_i$. Because impurities are relevant for ${\cal K}_e < 1$, for small enough temperatures the system flows to strong coupling (in the RG sense) and can be interpreted as a set of finite, decoupled 1D metallic segments; there is zero transmission across the impurity at $T=0$, and a power law behavior as a function of $T$. 

Under this simplistic picture, we expect two main effects on the noise behavior in the strong disorder regime. First, finite-size quantization effects for each segment will take place such that the energy level splitting of the segments will be, on average, $\Delta \varepsilon \sim \hbar v n_{\rm imp}$, with $n_{\rm imp}$ the impurity density. As such, if $\omega\lesssim\Delta\varepsilon$, then the probe will unlikely couple to the sample. More quantitatively, the probablity of finding a segment of size $\ell_i \geq v / \omega$ is exponentially small, $p\sim{\rm exp}(-n_{\rm imp}\ell_i)$. This analysis allows to define a minimum segment size in order to couple the spin probe to the sample. For instance, for sub-THz frequencies, we expect the minimun segment size to be on the order of $\ell_i \gtrsim v/\omega\sim 100\,{\rm nm}$ (here we used $v = 10^4\,{\rm m/s}$ and $\omega \lesssim 100\,{\rm GHz}$).

Second, because the distribution of lengths in each segment is expected to be random, we also expect highly inhomogeneous noise as a function of $x$. This is qualitatively different than the weak disorder case where, because the probe samples a large number of defects which are available within a distance R, noise is expected to be relatively homogeneous across $x$ due to self-averaging. 

Introducing an SU(2) degree of freedom leads to the same conclusions as the spin polarized state, namely, charge and spin density waves are perfectly reflected at the impirity for any value of repulsive interactions (for small enough $T$). Interestingly, if one breaks the SU(2) spin symmetry but preserves a spin U(1), then mixed phase are possible in which charge density waves are perfectly reflected and spin density waves are perfectly transmitted (or viceversa).\cite{1992kanefisher1} By separating charge and spin fluctuations, this peculiar behavior can potentially be detected with spin probes. 

\section{Beyond the minimal two channel model}
\label{sec:extension}

One dimensional systems can feature physics beyond our minimal two channel model. The simplest extension to our model is adding orbital degrees of freedom, for instance ladders or carbon nanotubes. Extension of our formalism to these cases follow the same lines as in the SU(2) spin degree of freedom case, i.e. increasing the number of fields, all of which satisfy Eq.(\ref{eq:action1d}) separately but with orbital-dependent Luttinger parameters. The large range of possibilities available for coupling the degrees of freedom via scattering or interactions give rise to a wide range of regimes in which noise can be dominated by different types of fluctuations, e.g. superconducting or ferromagentic fluctuations. 

In addition, although our attention was mainly on the qualitative features of noise, 1D physics is sensitive to microscopic details. It is often the case that several microscopic models explain, up to some degree, some experimental observation. Two notable examples are non-quantized conductivity for quantum spin Hall states, and the case of the 0.7 anomaly observed in conductance measurements in quantum wires.\cite{199607anomaly,199607anomaly2,201107anomalyreview} In the former case, several backscattering mechanism have been proposed for quantum spin Hall states, such as trapping of electrons in quantum dots, or disordered Rashba coupling.\cite{2009qshkondo,2010qshrashba,2011qshimpurity,2012qshbackscattering,2013helicalresistance,2014helicalglazman} In the latter case, two opposite pictures, namely a transition into a 1D Wigner crystal and the formation of a Kondo impurity, have been proposed to explain the data.  By exploting {\it both} the spatial and spectral (via $T$) resolution of spin probes, it may be possible to shed light on the operating mechanisms in these two intresting and somewhat controversial cases. 

In addition, low energy excitations may exhibit quasi-1D behavior. Such is the case of edges in the quantum Hall regime and edge magnetoplasmon, wherein charge density fluctuations at the edge induce not only edge currents but also bulk currents.\cite{1988magnetoplasmontheory} While the above discussion remains valid, it is now necessary to account for lateral currents (towards the bulk) in addition to the already studied edge currents. While not pursued here, an intriguing possibility is using spin probes to map the conducting channels in quantum Hall systems. 

\section{Summary}
\label{sec:conclusion}

The ability to sample charge and spin fluctuations in a wide range of lengthscales render spin probes an invaluable probing technique of 1D systems, particularly when the coupling between charge and spin modes is important. We outlined protocols which exploit the spin degree of freedom of the probe to measure charge and spin fluctuations in a wide range of 1D systems. Furthermore, we discussed the effects of scattering, interactions and internal structures of 1D carriers on temperature and probe-to-sample distance dependence of noise. We showed that these features can be accessed using readily availabe NV-based diamond probes. In the same spirit, spin probes are also promising candidates to explore a whole zoo of phenomena in 1D systems, such as Kondo impurities and ladders, thus opening intriguing new pathways to access charge and spin fluctuations in general 1D systems. 

\section{Acknowledgements}

We thank J. Sanchez-Yamagishi, T. Andersen, C. Bolech, V. Kasper, R. Schmidt, and S. Chatterjee for valuable insights and discussions. J.F.R.-N., and E.D. acknowledge support from the Harvard-MIT CUA, NSF Grant No. DMR-1308435, and AFOSR-MURI: Photonic Quantum Matter, award FA95501610323. K.~A. acknowledges support from the DOE-BES Grant No. DE-SC0002140, and the U.K. Foundation. T.~G. acknowledges support by the Swiss National Science Foundation under Division II. B.I.H. acknowledges support from the STC Center for Integrated Quantum Materials, NSF grant DMR-1231319. M.D.L. acknowledges support from the Harvard-MIT CUA, the Vannevar Bush Fellowship, and the Moore Foundation.


\appendix

\section{Relaxation rate of the spin probe}
\label{app:relaxationtime}

\renewcommand{\thefigure}{S\arabic{figure}}
\renewcommand{\theequation}{A\arabic{equation}}
\setcounter{equation}{0}
\setcounter{figure}{0}

There are two important protocols used to study magnetometry in solid-state system. In one approach, static magnetic fields can be measured by determining the Zeeman splitting of the spin probe. This approach, which allows to observe magnetic textures, has been used for single spin imaging,\cite{2013singlespinimaging} and domain walls.\cite{2016nvdomainwalls} In the other approach, which is the one discussed in the present work, the relaxation time of a spin-probe prepared in a pure state is measured. The relaxation time is governed by the time-dependent Hamiltonian ${\cal H}_{\rm spin} = (\hbar\omega/2)\sigma_z + g\mu_{\rm B}{\boldsymbol\sigma}\cdot{\bm B}(t)$, where we assume a spin-1/2 probe with an intrinsic level splitting $\hbar\omega$. Without loss of generality, we assume that the intrinsic polarizing field is in the $\hat{z}$ direction [here $g$ is the probe $g$-factor, $\mu_{\rm B}$ the Bohr magneton, and ${\bm B}(t)$ the wire-induced magnetic field]. We also assume that the 1D system is in thermal equilibrium, described by the density matrix $\rho_{\rm 1D} = \sum_n e^{-\varepsilon_n/k_{\rm B}T}|n\rangle\langle n|$. The absorption rate, $R_{\rm abs}$, and emission rate, $R_{\rm em}$, is obtained from Fermi Golden's rule using the initial state $|i\rangle = |-\rangle\otimes\rho_{\rm 1D}$ and $|i\rangle = |+\rangle\otimes\rho_{\rm 1D}$:
\begin{widetext}
\be
R_{\rm abs,em} = 2\pi \sum_{n,m} e^{-\varepsilon_n/k_{\rm B}T} \left[ {B}_{nm}^x{B}_{mn}^x +  {B}_{nm}^y{B}_{mn}^y \mp i {B}_{nm}^x{B}_{mn}^y \pm i {B}_{nm}^y{B}_{mn}^x\right]\delta (\omega \mp \omega_{nm}),
\ee
Here we used $\hbar = 1$, $g\mu_{\rm B}B_i $ was absorbed into $B_i$, ${B}_{nm}^i$ denotes $\langle n | {B}_i | m \rangle$, and ${\omega}_{mn}$ is the energy difference between states $n$ and $m$, $\omega_{nm} = \omega_n - \omega_m$. The relaxation rate is defined as $1/T_{1} = [R_{\rm abs} + R_{\rm em}]/2$. Explicitly, $1/T_{1}$ takes the form
\be
\begin{array}{rl}
1/T_{1}  = & \displaystyle\pi \sum_{n,m} e^{-\varepsilon_n/k_{\rm B}T} \left\{ {B}_{nm}^x{B}_{mn}^x \left[\delta (\omega + \omega_{nm})+\delta (\omega - \omega_{nm})\right] +  {B}_{nm}^y{B}_{mn}^y \left[\delta (\omega + \omega_{nm})+\delta (\omega - \omega_{nm})\right]\right.   \\
& \displaystyle\left. - 2{\rm Im}\left[{ B}_{nm}^x{B}_{mn}^y\right]\delta(\omega+\omega_{nm}) - 2 {\rm Im}\left[ {B}_{nm}^y{B}_{mn}^x\right] \delta(\omega-\omega_{nm}) \right\}.
\end{array}
\label{eq:1/T1}
\ee
It is straight-forward to show that $1/T_{1,z}$ can be expressed in terms of the noise tensor ${\cal N}_{ij}(\omega)$: 
\be
{\cal N}_{ij} (\omega) = \frac{1}{2}\int_{-\infty}^{\infty} dt \langle \{ B_i(t),B_j(0)\}\rangle e^{i\omega t} = \pi \sum_{nm} \rho_n \left[ B_{nm}^iB_{mn}^j\delta(\omega+\omega_{nm}) + B_{nm}^jB_{mn}^i\delta(\omega-\omega_{nm}) \right]. 
\label{eq:noisetensor}
\ee
\end{widetext}
In particular, by direct comparison between Eq.(\ref{eq:1/T1}) and (\ref{eq:noisetensor}), we find 
\be
1/T_{1} = {\cal N}_{xx}(\omega) + {\cal N}_{yy}(\omega) - 2{\rm Im}\left[{\cal N}_{xy}(\omega)\right]. 
\label{eq:t1p}
\ee
Whereas the diagonal components of ${\cal N}_{ij}(\omega)$ are real for all $\omega$, the off-diagonal components of ${\cal N}_{i j}(\omega)$ are complex numbers. 

For calculation purposes, it is convenient to cast Eq.(\ref{eq:t1p}) in terms of retarded and advanced correlation functions. As such, we first express the noise tensor ${\cal N}_{ij}(\omega)$  in terms of the spectral density of the magnetic field, 
\be
\begin{array}{rl}
\displaystyle{\cal S}_{ij} (\omega) = & \displaystyle\frac{1}{2}\int_{-\infty}^{\infty} dt \langle \left[ B_i(t),B_j(0)\right]_-\rangle e^{i\omega t} \\ & \\
= & \pi \sum_{nm} \rho_n \left[ B_{nm}^iB_{mn}^j\delta(\omega+\omega_{nm})\right. \\ & \\
& \left. - B_{nm}^jB_{mn}^i\delta(\omega-\omega_{nm}) \right],
\end{array}
\ee
via the fluctuation-dissipation theorem:
\be
{\cal N}_{ij}(\omega) = {\rm coth}(\omega/2T){\cal S}_{ij}(\omega). 
\ee
The relaxation time in terms of the spectral density (with restored units) is given by 
\be
\begin{array}{rl}
1/T_{1} = &  (g\mu_{\rm B}/\hbar)^2 {\rm coth}(\hbar\omega/2k_{\rm B}T) \\ & \\ & \times \left[{\cal S}_{xx}(\omega) + {\cal S}_{yy}(\omega) - 2{\rm Im}\left[{\cal S}_{xy}(\omega)\right]\right].
\end{array}
\ee
Secondly, ${\cal S}_{ij}(\omega)$ can be related to retarded and advanced correlators as ${\cal S}_{ij}(\omega) = -(i/2)\left[ {\cal C}_{B_iB_j}^{\rm R} (\omega) - {\cal C}_{B_iB_j}^{\rm A} (\omega)\right]$, where we denote ${\cal C}_{AB}^{\rm R,A}(\omega)  = \mp i \int_{-\infty}^{\infty} dt \Theta(\pm t) \langle \left[ A(t),B(0)\right]_-\rangle e^{i\omega t}$. 

\section{Wire-induced electromagnetic modes}
\label{app:emmodes}

\renewcommand{\theequation}{B\arabic{equation}}
\setcounter{equation}{0}

Here we find the electromagnetic modes induced by charge and spin density modes in a 1D system. Following the convention in the main text, the wire is aligned in the $\hat{\bm x}$ axis, and ${\bm r}_{\perp} = (y,z)$ are the coordinates transverse to the wire. Without loss of generality, we also assume that the probe is in position ${\bm r} = (0,0,R)$. In the following subsections, we first find the eigenfunctions ${G}_{m}^\mu(q,{\bm r}_\perp,\omega)$ associated to the electromagnetic field 
\be
A^\mu({\bm r},t) = \frac{1}{\sqrt{L}}\sum_{q\omega m}G_{m}^\mu(q,{\bm r}_\perp,\omega)e^{i(qx-\omega t)}\rho_m(q,\omega),
\ee
for charge ($m=e$) and spin ($m=x,y,z$) modes, and then compute ${\bm H}_m(q,{\bm r}_\perp,\omega)$ in Eq.(\ref{eq:bfield}) by taking the curl of ${G}_{m}^\mu(q,{\bm r}_\perp,\omega)$. 

\subsection{Charge-induced electromagnetic modes}

The electromagnetic eigenfunction associated with the 1D charge density is given by the solution of 
\be
\begin{array}{c}
\left[(\omega/c)^2 - q^2 + \nabla_{{\bm r}_\perp}^2 \right] G_{e}^\mu (q,{\bm r}_\perp,\omega) = \delta({\bm r}_\perp) d^\mu(q,\omega), \\ \\ d^\mu(q,\omega) = (1,\omega/q,0,0).
\end{array}
\label{eq:chargeevq}
\ee
Here we focus on evanescent wave solutions, $q \ge \omega/c$, because typical excitation wavevectors $q$ are on the order of $q \sim \omega/v_{\rm F}$, with $v_{\rm F}\ll c$. The 4-vector $d^\mu(q,\omega)$ originates the continuity equation, $\partial_t \rho_e + \partial_z j_e = 0$. The explicit solution of Eq.(\ref{eq:chargeevq}) is 
\be
G_{e}^\mu (q,{\bm r}_\perp,\omega) = - [d^\mu(q,\omega)/2\pi] K_0(\lambda|{\bm r}_\perp|), 
\ee
where $K_n$ denotes the $n$-th modified Bessel function of the second kind, and $\lambda = \sqrt{q^2 - (\omega/c)^2}$. The magnetic field at the position of the probe ${\bm r}=(0,0,R)$ is given by $\delta {\bm B} = \frac{1}{\sqrt{L}} \sum_{q \omega m} {\bm H}_{m}(q,{\bm r}_\perp,\omega) e^{i(qx - \omega t)} {\rho}_{m}(q,\omega) $, where ${\bm H}_{m}(q,{\bm r}_\perp,\omega)$ is found by taking the curl of Eq.(\ref{eq:chargeevq}): 
\be
{\bm H}_{e} = -\frac{1}{2\pi}\left( \begin{array}{c} 0 \\ (\omega\lambda/q) K_1(\lambda R) \\ 0 \end{array} \right) \approx -\frac{1}{2\pi}\left( \begin{array}{c} 0 \\ \omega K_1(q R) \\ 0 \end{array} \right).
\label{eq:bcharge}
\ee

\subsection{Spin-induced electromagnetic modes}

Similarly, the electromagnetic modes corresponding to the spin source are given by the solution of:
\be
[(\omega/c)^2 - q^2 + \nabla_{{\bm r}_\perp}^2] G_{m}^{\mu}(q,{\bm r}_\perp,\omega) = [0,\nabla\times(\delta({\bm r}_\perp)\hat{\bm e}_m)], 
\label{eq:spinevq}
\ee
where $\nabla$ reads $\nabla=(iq,\partial_y,\partial_z)$. The explicit solutions of Eq.(\ref{eq:spinevq}) are
\begin{widetext}
\be
G_x^\mu = \frac{1}{2\pi}\left( \begin{array}{c} 0 \\ 0 \\ \lambda K_1(\lambda r)\sin\theta \\ -\lambda K_1(\lambda r)\cos\theta \end{array} \right),
G_y^\mu = \frac{1}{2\pi}\left( \begin{array}{c} 0 \\ -\lambda K_1(\lambda r)\sin\theta \\ 0 \\ iqK_0(\lambda r) \end{array} \right),
G_z^\mu = \frac{1}{2\pi}\left( \begin{array}{c} 0 \\ -\lambda K_1(\lambda r)\cos\theta \\ -iqK_0(\lambda r) \\ 0 \end{array} \right).
\label{eq:vectorpotentialspin}
\ee
By taking the curl of Eq.(\ref{eq:vectorpotentialspin}) and using $\omega/c\ll q$, we find ${\bm H}_{m}(q,{\bm r}_\perp,\omega)$ at ${\bm r} = (0,0,R)$ given by
\be
{\bm H}_{x} = \frac{q^2}{4\pi}\left( \begin{array}{c} -K_0(qR) \\ 0 \\ 2iK_1(qR) \end{array} \right) , 
{\bm H}_{y} = -\frac{q^2}{4\pi}\left( \begin{array}{c} 0 \\ K_0(qR) + K_2 (qR) \\ 0 \end{array} \right), 
{\bm H}_{z} = \frac{q^2}{4\pi}\left( \begin{array}{c} -2i K_1(qR) \\ 0 \\ K_0(qR)-K_2(qR) \end{array} \right),
\label{eq:bspin}
\ee
\end{widetext}

\subsection{Spin-polarized systems}

For systems in which spin is polarized, we denote $\hat{\bm n} =(n_x,n_y,n_z)$ as the direction of polarization. The electromagnetic mode corresponding to a spin polarized source can be obtained from 
\be
[(\omega/c)^2 - q^2 + \nabla_{{\bm r}_\perp}^2]G_{\sigma}^{\mu}(q,{\bm r}_\perp,\omega) = \left[0,\nabla\times[\delta({\bm r}_\perp)\hat{\bm n}]\right].
\label{eq:spinpolevq}
\ee
The function $G_{\sigma}^\mu(q,{\bm r}_\perp,\omega)$ can be expressed in terms of $G_{m}^\mu(q,{\bm r}_\perp,\omega)$ in Eq.(\ref{eq:vectorpotentialspin}): $G_{\sigma}^\mu (q,{\bm r}_\perp,\omega)= \sum_{m=x,y,z} \hat{n}_mG_{m}^\mu(q,{\bm r}_\perp,\omega)$. Similarly, the ${\bm H}_{\sigma}$ components of the magnetic field is ${\bm H}_{\sigma}(q,{\bm r}_\perp,\omega) = \sum_{m = x,y,z} \hat{n}_m{\bm H}_{m}(q,{\bm r}_\perp,\omega)$. 

\section{Relaxation time and density-density correlators}
\label{app:effectiveaction}

\renewcommand{\theequation}{C\arabic{equation}}
\setcounter{equation}{0}

In this section we show the connection between magnetic field fluctuations and charge/spin density fluctuations in the wire [c.f. Eq.(\ref{eq:noise2})]. To evaluate the anticommutator in Eq.(\ref{eq:noise}) which gives rise to spin relaxation, we first use the fluctuation-dissipation theorem 
\be
\begin{array}{r}
\int_{-\infty}^{\infty}dt \langle \{\delta B_i(t) ,\delta B_j(0)\} \rangle e^{i\omega t} = {\rm coth}(\hbar\omega / 2 k_{\rm B} T) \\ \\ \int_{-\infty}^{\infty}dt \langle [\delta B_i(t) ,\delta B_j(0)] \rangle e^{i\omega t}. 
\end{array}
\ee
The right-hand side of this equation can be written in terms of retarded and advanced correlation functions, 
\be
\begin{array}{l}
\int_{-\infty}^{\infty} dt \langle[\{\delta B_i(t),\delta B_j(0)\}]\rangle e^{i\omega t}=  \\ \\
\qquad\qquad{\cal C}_{\delta B_i \bar{\delta B_j}}^{\rm R}(q,\omega)-{\cal C}_{\delta B_i \bar{\delta B_j}}^{\rm A}(q,\omega).
\end{array}
\ee 
Expressing $\delta B_i $ in terms of the (orthogonal) electromagnetic modes, Eqs.(\ref{eq:bcharge}) and (\ref{eq:bspin}), and using the separation between charge and spin fluctuations, we find 
\be
\begin{array}{c}
\int_{-\infty}^{\infty}dt \langle \{\delta B_i(t) ,\delta B_j(0)\} \rangle e^{i\omega t} = \sum_{qm} A_{ij}^{m}(q,{\bm r}_\perp,\omega) \\
 \left[{\cal C}_{\rho_m\bar{\rho}_m}^{\rm R}(q,\omega) - {\cal C}_{\rho_m\bar{\rho}_m}^{\rm A}(q,\omega)\right],  
\end{array}
\label{eq:mainnoise}
\ee
where 
\be
\begin{array}{rl}
A_{ij}^{m} (q,{\bm r}_\perp,\omega) =  & {\rm coth}(\hbar\omega / 2 k_{\rm B} T)  \\ & \\ & \times H_{m,i}(q,{\bm r}_\perp,\omega)\bar{H}_{m,j}(q,{\bm r}_\perp,\omega).
\end{array}
\ee
Using ${\cal C}_{\rho_m\bar{\rho}_m}^{\rm A}(q,\omega) = [{\cal C}_{\rho_m\bar{\rho}_m}^{\rm R}(q,\omega)]^* $ and multiplying Eq.(\ref{eq:mainnoise}) by $\frac{1}{2}(g_{\rm s}\mu_{\rm B}/\hbar)^2$, we obtain Eq.(\ref{eq:noise2}). 

\section{Memory function for disorder}
\label{app:memoryfunction}

\renewcommand{\theequation}{D\arabic{equation}}
\setcounter{equation}{0}

For the SU(2) case, the memory functions are defined in terms of the retarded correlator ${\cal C}_{f_mf_m}^{\rm R}(q,\omega)$, where $f_m(x) = [\Pi_m(x),\int dx'{\cal H}_{\rm dis}(x')]$ captures the momentum relaxation rate of $\Pi_m$. Explicitly, $f_m(x)$ takes values $f_m(x) = (2v_m{\cal K}_m / a) u(x) e^{i\sqrt{2}\phi_e(x)}\cos[\sqrt{2}\phi_\sigma(x)]+{\rm h.c.}$. Calculations of correlations functions where the field $\phi_m(x)$ appears in the exponent is straight-forward but tedious. A detailed step-by-step procedure is discussed in Appendix C of Ref.[\onlinecite{giamarchibook}]. Using $\langle u(x) \bar{u}(x')\rangle = D\delta(x-x')$, the correlation function can be expressed as in Eq.(\ref{eq:memoryfunction}) with parameters:
\be
\begin{array}{l}
\Gamma_m = (2\pi)^{{\cal K}_{\rm t} +1}D{\cal K}_m^2 v_m^{{\cal K}_{\rm t}}/v_e^{{\cal K}_e}v_\sigma^{{\cal K}_\sigma}, \\ \\
\displaystyle \alpha_{m} ={\cal K}_t -2, \\ \\ 
\displaystyle {\cal F}_{m}(x)= \sin(\pi {\cal K}_t/2) \\
\displaystyle\quad\quad\times\frac{B({\cal K}_t/2 - ix,1-{\cal K}_t)-B({\cal K}_t/2 , 1-{\cal K}_t)}{x}, 
\end{array}
\ee
where $B(x,y)$ is the Beta function and ${\cal K}_t = {\cal K}_e + {\cal K}_\sigma$. For repulsive interactions, our numerical estimates show that that the dimensionless function ${\cal F}_{m}(x)$ can be approximated as ${\cal F}_{m}(x\lesssim 1) \approx \beta i$, where $1 \lesssim \beta \lesssim 5$ for a wide range wide range of ${\cal K}_{e,\sigma}$ and $x$ values. 

The same procedure holds for the scattering potential in Eq.(\ref{eq:disordersp}), where $f(x) = (v_e{\cal K}_e / a) u_n(x) e^{i2n\phi_e(x)}+{\rm h.c.}$. Assuming an uncorrelated scattering potential, $\langle u_n(x)u_n(x') \rangle = D_n \delta(x-x')$ leads to Eq.(\ref{eq:memoryfunction}) with parameters:
\be
\begin{array}{c}
\displaystyle{\Gamma}_n = \frac{2^{2 n {\cal K}_e - 2}}{\pi}\frac{D_n {\cal K}_e}{v_e}, \quad 
 \alpha_n =2n{\cal K}_e -2, \\ \\ 
\displaystyle {\cal F}_n(x)= \sin(\pi {\cal K}_e)\frac{B({\cal K}_e - ix,1-2{\cal K}_e)-B({\cal K}_e , 1-2{\cal K}_e)}{x}, 
\end{array}
\ee
 For $n=1$ and repulsive interactions, our numerical estimates show that that the function ${\cal F}_n(x \lesssim 1)$ can be approximated as $f_1(x) \approx \beta_1 i$, where $1 \lesssim \beta_1 \lesssim 3$ for a wide range wide range of ${\cal K}_e$ and $x$ values. For $n=2$, the function ${\cal F}_n(x)$ can be approximated as ${\cal F}_2( x \lesssim 1) \approx \beta_2 i$, where $0.1 \lesssim\beta_2\lesssim 1$ for a wide range wide range of ${\cal K}_e$ and $x$ values. 


\end{document}